\documentclass[aps,pra,twocolumn,groupedaddress,superscriptaddress,bibnotes,amsfonts,
citeautoscript,a4paper]{revtex4-1}
\usepackage{latexsym}
\usepackage{amsmath}
\usepackage{amssymb}
\usepackage{graphicx}
\usepackage[T1]{fontenc}
\usepackage[open]{bookmark}
\usepackage{hyperref}
\hypersetup{colorlinks=true,allcolors=blue}

\begin{document}

\title{Reconfigurable chirality with achiral excitonic materials in the strong-coupling regime}

\author{P. Elli Stamatopoulou}
\affiliation{Center for Nano Optics, University of Southern Denmark,
Campusvej 55, DK-5230 Odense M, Denmark}
\author{Sotiris Droulias}
\affiliation{Department of Digital Systems, University of Piraeus, GR-18534, Piraeus, Greece}
\author{Guillermo P. Acuna}
\affiliation{Department of Physics, University of Fribourg, Chemin du Mus\'{e}e 3, Fribourg CH-1700, Switzerland}
\author{N. Asger Mortensen}
\affiliation{Center for Nano Optics, University of Southern Denmark,
Campusvej 55, DK-5230 Odense M, Denmark}
\affiliation{Danish Institute for Advanced Study, University of Southern Denmark,
Campusvej 55, DK-5230 Odense M, Denmark}
\author{Christos Tserkezis}
\affiliation{Center for Nano Optics, University of Southern Denmark,
Campusvej 55, DK-5230 Odense M, Denmark}

\date{\today}

\begin{abstract}
We introduce and theoretically analyze the concept of manipulating
optical chirality via strong coupling of the optical modes of chiral
nanostructures with excitonic transitions in molecular layers or
semiconductors. With chirality being omnipresent in chemistry and
biomedicine, and highly desirable for technological applications
related to efficient light manipulation, the design of nanophotonic
architectures that sense the handedness of molecules or generate
the desired light polarization in an externally controllable
manner is of major interdisciplinary importance. Here we propose
that such capabilities can be provided by the mode splitting resulting
from polaritonic hybridization. Starting with an object with well-known
chiroptical response ---here, for a proof of concept, a chiral sphere---
we show that strong coupling with a nearby excitonic material generates
two distinct frequency regions that retain the object's chirality
density and handedness, which manifest most clearly through
anticrossings in circular-dichroism or differential-scattering
dispersion diagrams. These windows can be controlled by the
intrinsic properties of the excitonic layer and the strength of
the interaction, enabling thus the post-fabrication manipulation
of optical chirality. Our findings are further verified via
simulations of the circular dichroism of a realistic chiral
architecture, namely a helical assembly of plasmonic nanospheres
embedded in a resonant matrix.
\end{abstract}

\maketitle

\section{Introduction}
Chirality is encountered widely in nature, describing, in
its most broad definition, the possibility of an object not
being superimposable onto its mirror image. It is a common
characteristic of biomolecules, massively determining their
chemical function and interactions. For instance, chiral
assemblies have been shown to regulate autophagy ---the
mechanism responsible for recycling unnecessary components
within cells--- whose abnormal regulation is associated with
diseases such as cancer and diabetes~\cite{sun_natcom9}. In
the pharmaceutical and food industry opposite handedness is
oftentimes the factor that distinguishes between harmful and
beneficial substances~\cite{nguyen_ijbs2,alvarez_trac123}.
It is, therefore, not to wonder why sensing and controlling
chirality has remained the subject of intense study for
decades, with optical characterization techniques attracting
particular attention~\cite{Berova_Wiley2000}. Besides
biomedical interest, chiral structures have attracted
the attention of the photonic  ---particularly
plasmonic~\cite{hentschel_sciadv3}--- community, due to
their ability to distinguish between left-/right-handed
circularly polarized (L/RCP) light, rendering them ideal
platforms for the study of optical activity and the
chiroptical effects associated with it~\cite{petronijevic_omex12}.
They have been proposed as building blocks for metamaterials
for the generation of superchiral light~\cite{collins_aom5},
negative refraction~\cite{plum_prb79} and as broadband
circular polarizers~\cite{gansel_sci352}. Control of
chiral matter--light interactions at the quantum level
has also received significant attention for applications
in quantum information technologies~\cite{lodahl_nat541,aiello_nn16}.

The origin of optical activity in photonic systems can be
attributed to different manifestations of chirality. Small metal
clusters have been shown to exhibit intrinsic chirality, owing
to the asymmetric arrangement of achiral adsorbates on their
surfaces~\cite{dolamic_natcom3}. Moreover, nanoparticles (NPs) ---most
commonly metallic--- arranged in specific geometries can
demonstrate optical chirality stemming from their structural
characteristics~\cite{lan_jacs137}. Finally, a chiral response
can also be induced externally, either by means of chiral
inclusions in achiral metasurfaces~\cite{droulias_prb102}
or by external magnetic fields~\cite{maccaferri_jap127}. For
example, magnetic circular dichroism (MCD) describes a special
case of magnetically induced chirality, where an otherwise
achiral object becomes optically active in the presence of
a static magnetic field, as observed in gyroelectric
media~\cite{caridad_prl126,stamatopoulou_prb102}.

Chirality is typically quantified by measuring the circular
dichroism (CD) and the differential scattering (DS) of a
sample~\cite{Barron_Cambridge2004}, referring to the different
absorption and scattering, respectively, of LCP and RCP light
that it undergoes. However, conventional CD spectroscopy is
restricted to a narrow range of applications, due to the
general weakness of the corresponding signals in natural
chiral materials~\cite{warning_nn15}. Several approaches
for enhancing the sensitivity of the measurements have
been proposed, from superchiral fields~\cite{hendry_natnano5}
and metamaterials~\cite{droulias_nl20}, to resonant
single NPs and NP arrays~\cite{valev_admat25,mu_ns13,etxarri_prb87,lasa_acsp7,graf_acsp6}
or self-assemblies~\cite{kuzyk_nat483}. Plasmon-induced
chirality has been demonstrated both theoretically~\cite{govorov_jpcc115}
and experimentally~\cite{maoz_jacs134,duan_ns7} through
interaction of chiral molecules with achiral metallic NPs,
which are famous for strongly enhancing the local
electromagnetic (EM) field. At the same time, considerable
effort has also focused on high-index dielectric NPs and
metasurfaces~\cite{solomon_acsphot6,raziman_acsphot6,zhang_ns17,yao_ns10}.

Here we propose an efficient route for manipulating the
response of chiral structures, by means of reconfiguring their
photonic environment. Excitonic materials that sustain resonant
modes stemming from excitonic transitions at the atomic level
have already been proposed and explored as a means to control
the optical response in conventional plasmonic assemblies.
In recent works, the marriage of plasmonic NPs with
\emph{J}--aggregates of organic molecules or with
two-dimensional (2D) transition-metal dichalcogenides (TMDs)
that exhibit excitonic transitions, operating in the strong
coupling regime, has paved the way for a new era of flexible
and multifunctional platforms for modern nanophotonic
applications~\cite{fofang_nl10}. Such composite
architectures give rise to hybrid states, termed exciton
polaritons or plexcitons, that combine the properties of
both light and matter~\cite{torma_rpp78}. 
In addition to plexcitons, high-index
dielectrics are more recently being considered as alternative
environments for excitonic strong coupling, potentially as a
way to surpass the hurdle of high Ohmic losses dominating
metals, or to design novel polaritons with a magnetic
character~\cite{tserkezis_prb98,todisco_nanoph9,castellanos_acsphot7}.
Strong coupling of a chiral excitonic material with an achiral
resonator (e.g. a plasmonic NP) has already been shown to
generate wide anticrossings in experimentally measured CD spectra~\cite{wu_nn15}.
Here we theoretically explore chiral structures whose CD and DS
is rendered reconfigurable via the addition of an achiral
excitonic environment. Such a configuration has been studied
in the case of two coupled metallic nanorods~\cite{zhu_nl21},
but it is much more general and can be observed in any chiral
system, as we discuss below. As a proof or principle, we first
probe a simple, analytically solvable configuration, that of
a non-dispersive high-index dielectric sphere with intrinsic
chirality, covered with an excitonic layer. Then we proceed
to a system with structural chirality, namely silver (Ag)
NPs in a helical arrangement. We observe the distinct
behaviour of the system when operating in the strong
coupling regime, dominated by large spectral anticrossings
not only in absorption spectra but also in CD,
showcasing thus the engineering possibilities offered by
these hybrid systems.

\section{Results and Discussion}

The chirality of a system is encoded within its constitutive
relations, that relate the electric field $\mathbf{E}$ and
magnetic induction $\mathbf{B}$ to the displacement field $\mathbf{D}$
and the magnetic field $\mathbf{H}$ via the introduction of the
Pasteur parameter $\kappa$~\cite{condon_rmp9},
\begin{subequations}
\begin{gather}
\mathbf{D} = 
\varepsilon \varepsilon_{0} \mathbf{E} +
i \, (\kappa/c) \mathbf{H} \\ 
\mathbf{B} = 
\mu \mu_{0 }\mathbf{H} -
i \, (\kappa/c) \mathbf{E},   
\end{gather}\label{eq:const_rel}
\end{subequations}
which typically takes complex values. In eqn~(\ref{eq:const_rel}),
$\varepsilon$ and $\mu$ denote the relative permittivity and
permeability of the chiral medium respectively, $\varepsilon_{0}$
and $\mu_{0}$ the corresponding vacuum values, and
$c = 1/ \sqrt{\varepsilon_{0} \mu_{0}}$ is the speed of light
in vacuum. The eigenmodes of an infinite isotropic such medium are
L/RCP ($+/-$) waves that propagate with refractive indices
$n_{\pm} = n_{\rm{c}} \pm \kappa$ respectively, where
$(n_{+} + n_{-})/2 = n_{\rm{c}} = \sqrt{\epsilon \mu}$ is
the average (background) refractive index. The Pasteur parameter
$\kappa$ describes coupling between electric and magnetic field
components, and is generally related to the $\textbf{D} \cdot
\textbf{B}$ product. 

As a means to quantify the chiral properties
of an EM field, the local chirality density is employed, defined
as~\cite{tang_prl104}
\begin{equation}\label{eq:chirdens}
C = -\frac{\omega}{2}\, 
\mathrm{Im} \{\mathbf{D}^{*} \cdot \mathbf{B}\},
\end{equation}
where $\omega$ is the angular frequency of light. 
From eqn~(\ref{eq:chirdens}) it is clear that the chirality
density is associated with parallel electric and magnetic
components that are properly phase shifted. 
Regarding the sign of chirality density, it is interesting to note that it is preserved when the duality symmetry of the EM field is fulfilled~\cite{etxarri_prb87,fernandez_prl111}.

\begin{figure}[ht]
\centering
\includegraphics[width=0.65\columnwidth]{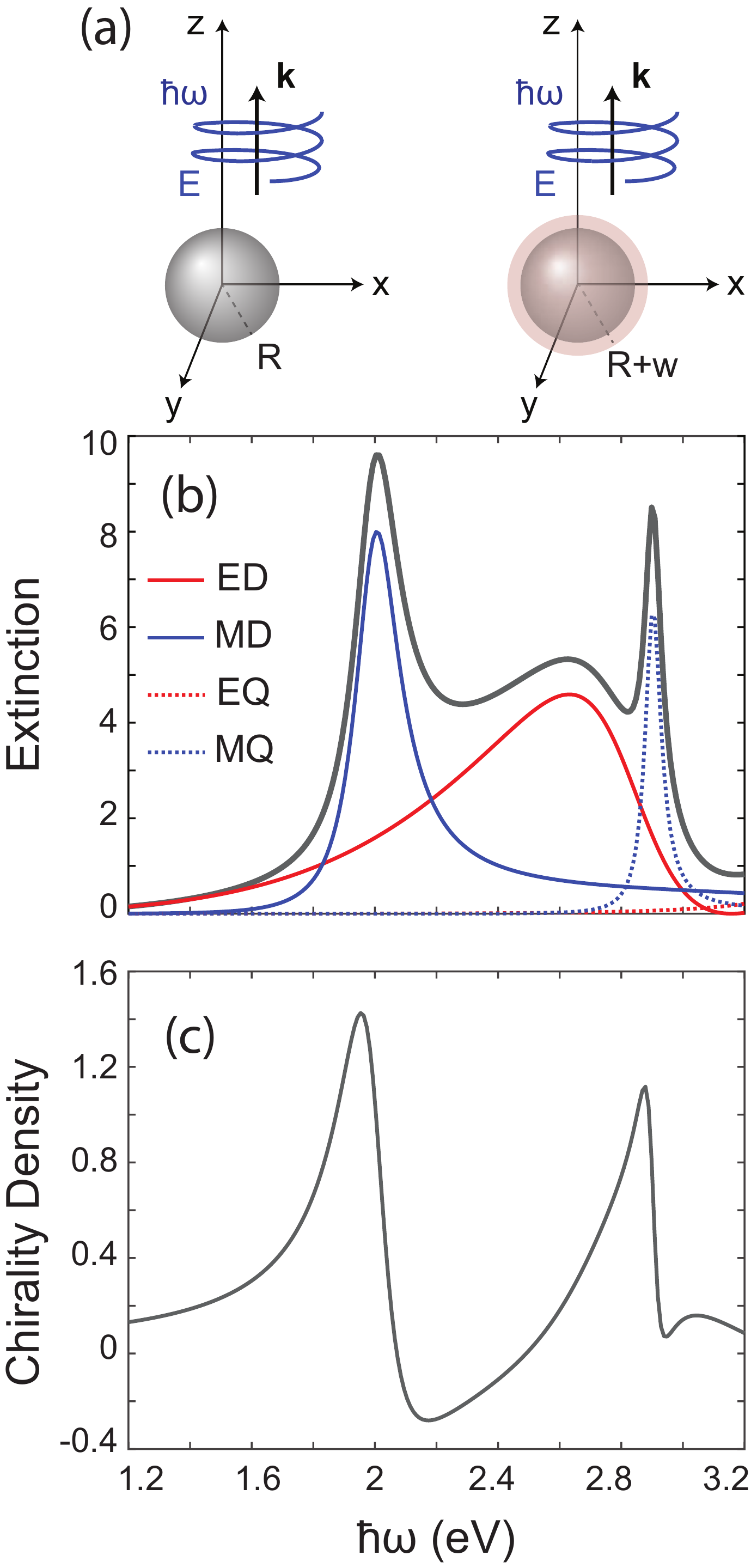}
\caption{
(a) Sketch of a chiral sphere illuminated by a circularly
polarized wave of energy $\hbar \omega$ and wavevector
$\mathbf{k}$ (left), and the corresponding exciton-covered
NP studied in Figs.~\ref{fig:achNP_coupled}
and~\ref{fig:chirNP_coupled}. 
(b) Extinction cross section (grey curve), normalized to
the geometrical cross section $\pi R^{2}$ of an achiral
sphere with radius $R = 85$\;nm and dielectric function
$\varepsilon = 12.1 + 0.001 i$, excited by an LCP wave.
The contribution from electric/magnetic dipolar/quadrupolar 
Mie modes is shown by red/blue solid/dotted lines.
(c) Chirality density of the same NP integrated over its
volume and normalized to the corresponding value of
chirality density in the bulk of the medium (see ESI).
}\label{fig:achNP}
\end{figure}

\begin{figure*}[ht]
\centering
\includegraphics[width=2.0\columnwidth]{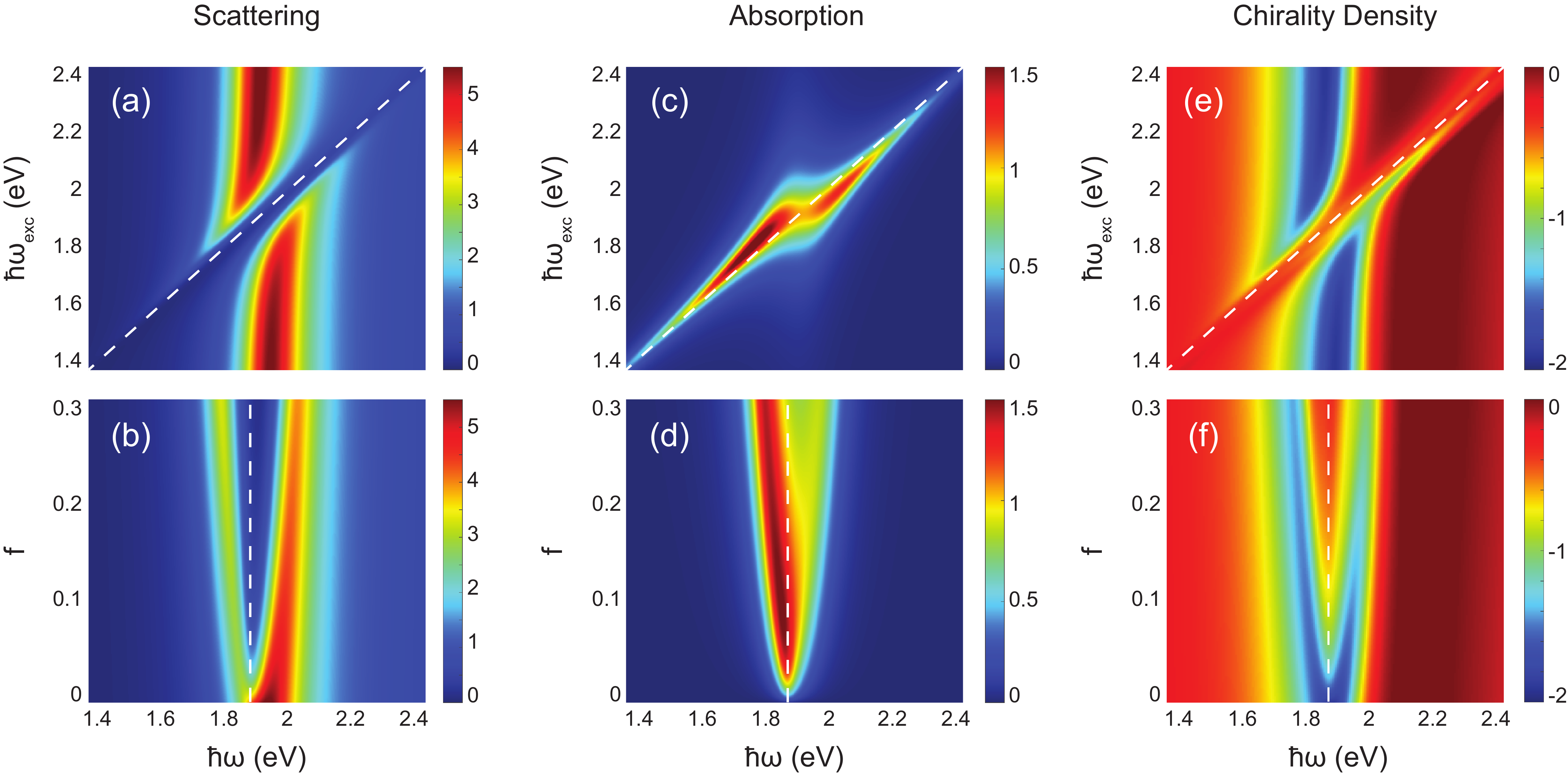}
\caption{ 
(a-b) Scattering and (c-d) absorption cross sections (normalized to the geometrical cross section) and (e-f) chirality density spectra (normalized to the corresponding bulk values)
of the achiral sphere of Fig.~\ref{fig:achNP}(b) coated with an excitonic
shell of thickness $w = 20$~nm as a function of (a,c,e) the excitonic transition
energy $\hbar \omega_{\mathrm{exc}}$ (for $f = 0.2$) and  (b,d,f) the excitonic
oscillator strength $f$ (for $\hbar\omega_{\mathrm{exc}} = 1.88$~eV). For the
excitonic material we use the parameters $\varepsilon_{b} = 3$ and 
$\hbar \gamma_{\mathrm{exc}} = 0.05$~eV. Here (and in Figs.~\ref{fig:chirNP_coupled} and \ref{fig5}), superimposed dashed white lines serve as
guides to the eye for tracing the energy of the excitonic resonance. In all panels the NP is excited by LCP light.}
\label{fig:achNP_coupled}
\end{figure*}

To explore the effect of an excitonic environment on the chirality response
of optically active matter, it is important to first understand the impact
of the excitonic coating on its resonant features, i.e. the overlapping
electric and magnetic modes. To this end, and in order to maintain complexity
at a minimum, we first study in Fig.~\ref{fig:achNP}(b) the optical response
of a spherical high-index, achiral ($\kappa = 0$) NP, $85$~nm in radius,
characterized by a constant permittivity $\varepsilon = 12.1 + 0.001 i$ and
permeability $\mu = 1$, illuminated by a plane EM wave [see left-hand 
sketch in Fig.~\ref{fig:achNP}(a)]. Due to the very small
imaginary part of the permittivity, the absorption in the NP is negligible and
light is mostly scattered~\cite{Bohren_Wiley1983}. Rigorous analytic Mie
theory~\cite{mie_annphys330} in Fig.~\ref{fig:achNP}(b) predicts a magnetic
dipolar (MD) mode at around $2$~eV, a broad electric dipolar mode (ED)
centered at about $2.6$~eV, and a sharp magnetic quadrupolar mode
(MQ) at $2.9$~eV, whereas higher-order modes (e.g. the electric quadrupole,
EQ) are negligible in the energy window we operate. 
The electric and
magnetic modes exhibit a significant spectral overlap in certain energy windows. 
We therefore expect local chirality density hotspots at these energies,
at the points that host parallel electric and magnetic components, 
not only in the space surrounding the sphere, but also in its volume, 
since the scattered fields and the ones developing inside the NP are resonant at similar energies; the Mie coefficients describing each, exhibit poles at the same eigenfrequencies.
Indeed, the chirality density spectrum integrated over the volume of the NP (see ESI for
the analytic expression) in Fig.~\ref{fig:achNP}(c) reveals local extrema close to the
energies where magnetic and electric modes intersect. We stress, however, that this overlap
should be taken as an indication, rather than a condition for high chirality density. After all, extinction is mostly due to the fields scattered in the environment of the sphere, rather than fields in its volume where the chirality density is evaluated.

Once the excitonic component is introduced, here as a concentric shell
of thickness $w$ [see right-hand sketch in Fig.~\ref{fig:achNP}(a)]
to retain the existence of exact analytic solutions, the system features
two types of resonant modes, the electromagnetic Mie resonances and the
matter-like excitonic resonances~\cite{tserkezis_prb98}. The dielectric
function of the excitonic material is for simplicity given by the Lorentz
model
\begin{equation}\label{eq:eq1}
\varepsilon (\omega) = 
\varepsilon_{b} - 
\frac{f \omega_{\mathrm{exc}}^{2}}
{\omega^{2} - \omega_{\mathrm{exc}}^{2} + 
i \gamma_{\mathrm{exc}} \omega},
\end{equation}
where $\varepsilon_{b}$ is the background permittivity, $\omega_{\mathrm{exc}}$
is the frequency of the excitonic transition (described in the model as a
harmonic oscillator), $f$ is the oscillator strength, and $\gamma_{\mathrm{exc}}$
the damping rate. When the resonant features of each constituent (Mie-resonant and excitonic) are tuned to
the same frequency, they couple in a manner analogous to the classical problem
of two coupled harmonic oscillators. If the interaction is strong enough, i.e.
in the strong-coupling regime, the energy states of the Lorentz oscillator and
the optical modes are tied together, leading to the emergence of two hybrid
light--matter resonances, identified in dispersion spectra by the avoided
crossing or anticrossing of the modes, also referred to as \emph{Rabi splitting} due to
its analogy to quantum optics~\cite{torma_rpp78,tserkezis_rpp83}.

Fig.~\ref{fig:achNP_coupled} shows the normalized (to the geometrical cross section) scattering and absorption cross
section of the composite NP (achiral NP and excitonic shell) for varying parameters
of the Lorentz oscillator, as well as the corresponding chirality density
of the coupled system. We note here that we chose to couple the exciton resonance to the MD mode of the particle, since it exhibits a well-defined linewidth, but similar coupling can be achieved with the ED and the higher order modes of the particle. Therefore, since the MD mode is the one primarily coupling to the excitonic transition, it is deliberately the only one taken into account in the Mie expansions in Figs.~\ref{fig:achNP_coupled}(a)--(d), in order to understand the resulting modification of the optical response of the particle on a fundamental level. In practice, however, there is a low contribution of the ED mode in the 1.4 -- 2.4~eV energy window, i.e. the low-energy tail of the red line in Fig.~\ref{fig:achNP}(b), and a weak coupling also occurs between the exciton and the ED mode.
The anticrossing trend between the MD mode and the excitonic resonance, as
the latter is shifted in energy, illustrated in the scattering spectrum of
Fig.~\ref{fig:achNP_coupled}(a), reveals that the system is possibly
strongly coupled. The width of the split serves as a qualitative measure
of the coupling strength and can be tuned via the oscillator strength of
the excitonic transition [see Fig.~\ref{fig:achNP_coupled}(b)]. In
Fig.~\ref{fig:achNP_coupled}(c) and (d) we show the effect of the
excitonic shell on the absorption spectrum for varying excitonic
transition energy and oscillator strength, respectively. The pattern
observed in Fig.~\ref{fig:achNP_coupled}(c) where the mode splitting
is not clearly discernible is sometimes referred to as \emph{induced
transparency}~\cite{antosiewicz_acsphot1}. Examination of the strong coupling criteria, however, reveals that the system indeed operates in the strong coupling regime (see ESI). The chirality density, only taking the dipolar modes (MD, ED) of the coupled structure into account, follows the anticrossing behaviour of the scattering
spectrum [see Fig.~\ref{fig:achNP_coupled}(e) and (f)]. Since the change in chirality density
is significant, we expect a highly tunable CD and DS signal once we
introduce chiral properties to the dielectric core of the structure.

\begin{figure}[t]
\centering
\includegraphics[width=\columnwidth]{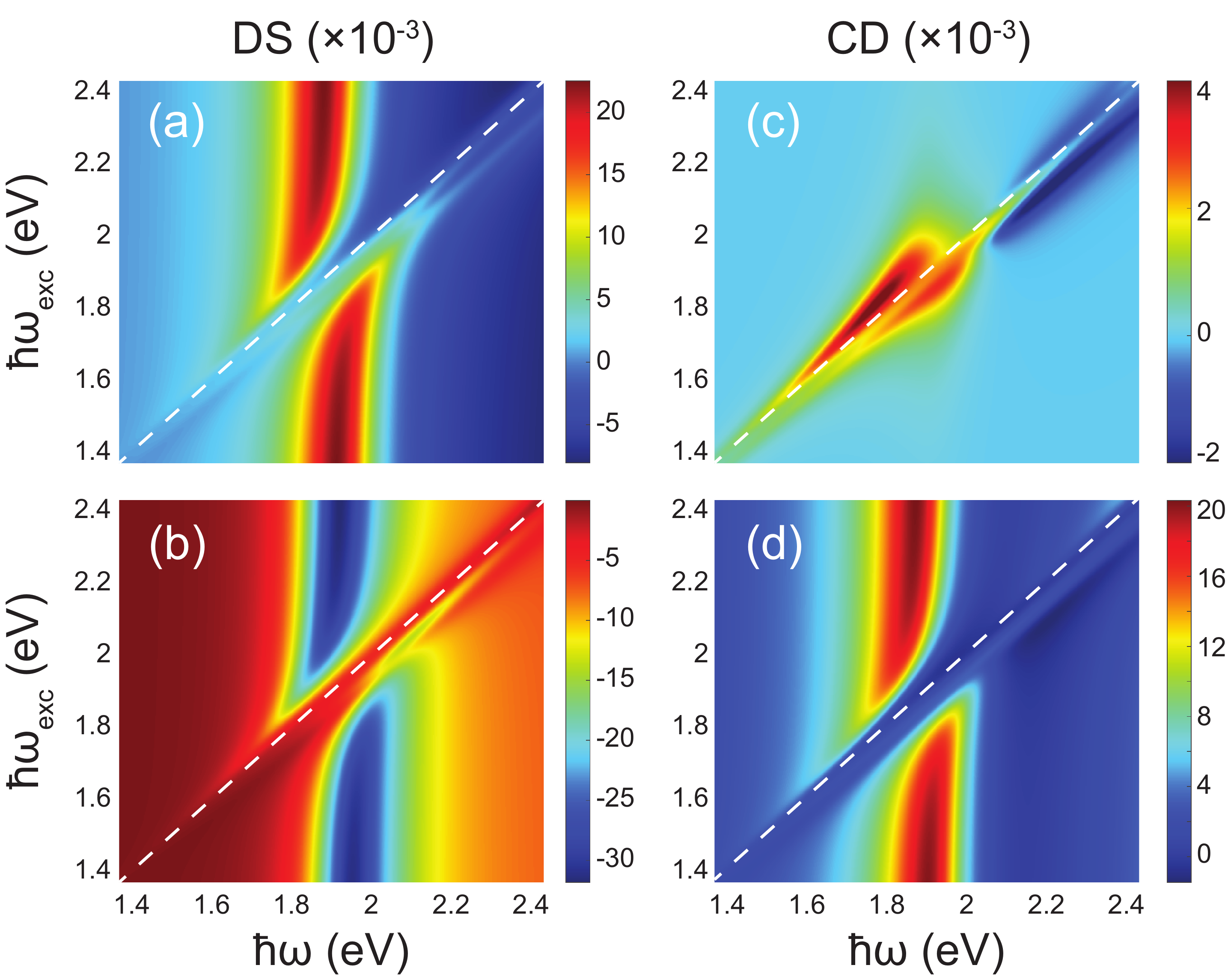}
\caption{
Photon versus exciton energy colour map of the (a-b) DS and (c-d) CD 
corresponding to the same coated NP as in
Fig.~\ref{fig:achNP_coupled}, here with a chiral core described
by the Pasteur parameter $\kappa = 0.001$ (panels a and c) and 
$\kappa = 0.001 i$ (panels b and d).
}\label{fig:chirNP_coupled}
\end{figure}

We next explore the impact of the excitonic environment on an optically
active configuration. Fig.~\ref{fig:chirNP_coupled}(a) and (c) show the
CD and DS of the same coated dielectric NP, now with a chiral core
characterized by $\kappa = 0.001$. This particular value of the
chirality parameter serves two purposes: on the one hand, it is small
enough to ensure that the electric and magnetic modes will be coupled
but will otherwise remain unaffected, and on the other hand it is
real-valued, leading to a typically very weak CD signal of the core
alone (see ESI). In passing, here we define the DS and CD signal as~\cite{Barron_Cambridge2004}
\begin{subequations}\label{Eq:CD-DS}
\begin{gather}
\mathrm{CD} = 
\sigma_{\mathrm{abs}}^{\mathrm{R}} -
\sigma_{\mathrm{abs}}^{\mathrm{L}} \\
\mathrm{DS} = 
\sigma_{\mathrm{scat}}^{\mathrm{R}} -
\sigma_{\mathrm{scat}}^{\mathrm{L}},
\end{gather}
\end{subequations}
where $\sigma_{\mathrm{abs/scat}}^{\mathrm{R/L}}$ is the absorption/scattering
cross section corresponding to an R/LCP incident light wave normalized to the
geometrical cross section, analytic expressions for which can be found in several textbooks for optically active media~\cite{Bohren_Wiley1983}.

As already mentioned, for a weak real-valued chiral parameter $\kappa$ one
would perhaps expect a negligible CD signal. On the contrary,
Fig.~\ref{fig:chirNP_coupled}(c) proves that the presence of the excitonic
material has led to a significant CD component, even comparable to the DS
component as shown in Fig.~\ref{fig:chirNP_coupled}(a). 
The reason is that chirality lifts the degeneracy of the Mie modes between excitation with LCP and RCP waves, as is already evident in the DS calculations. As a result, and despite the absence of absorbing chiral regions, asymmetric absorption takes place in the achiral regions of the combined shell-sphere system, mainly the excitonic shell, thus leading to nonzero CD signal. This observation is in accord with theoretical predictions of nonvanishing CD signals in achiral metasurfaces with chiral inclusions characterized by entirely real $\kappa$~\cite{droulias_prb102}. Additionally, this is the reason why, eventually, the CD signal inherits the behaviour of the absorption in the achiral particle, as presented in Fig.~\ref{fig:achNP_coupled}(c).
We note, that ---in contrast with Fig.~\ref{fig:achNP_coupled}--- what we plot in all panels of
Fig.~\ref{fig:chirNP_coupled} is the \emph{total} optical response of the
structure, including higher order electric and magnetic modes, since the MD mode, albeit the primary radiating channel, is not the only mode that is supported in the operating energy window. On the other hand, the effect of the
excitonic layer on the already significant DS component is the splitting of
the resonant spectral feature. Similar to the achiral structure, those
effects can become more prominent as we tune the oscillator strength of
the excitonic material (see ESI). 

Figs.~\ref{fig:chirNP_coupled}(b) and (d)
show the effect of the excitonic shell on the chiral properties of an NP
with imaginary $\kappa = 0.001i$, in which case the uncoated NP already
exhibits a significant CD signal (see ESI). Here, the DS spectrum
exhibits, as before, an anticrossing trend as in the case of real-valued
$\kappa$. The CD spectrum, however, attains a completely different behavior.
What we observe in this case is the splitting of the spectral feature as
the two components are detuned, similarly to the DS spectrum. 
When $\kappa$ is imaginary-valued, the contribution of the chiral material regions to asymmetric absorption dominates over that in the achiral regions. In this case, CD locally behaves as $ \mathrm{CD}(\textbf{r}) \propto \mathrm{Im} (\kappa) C (\textbf{r})$ \cite{etxarri_prb87}, and therefore inherits the anticrossing trend of the chirality density $C$, as seen in Fig.~\ref{fig:achNP_coupled}(e).

\begin{figure}[b!]
\centering
\includegraphics[width = 1.0\columnwidth]{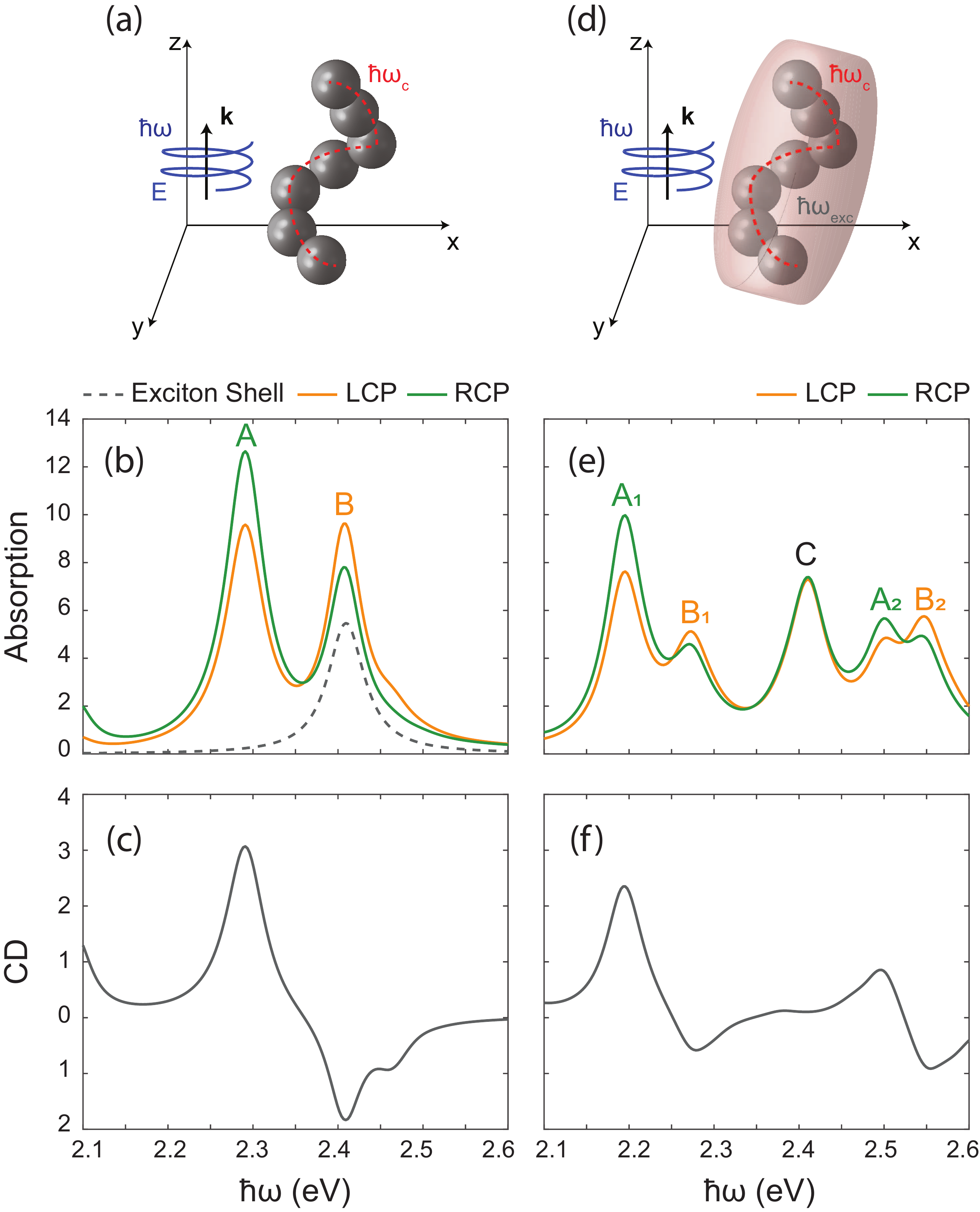}
\caption{
(a) Schematic of an Ag-NP helix: 7 nanospheres, of radius $R = 5$\;nm
each, revolve around a hypothetical supporting pillar along the $z$
axis, with steps of $30^{\circ}$, separated by a vertical distance
$2R$ along $z$. The helix is illuminated by circularly polarized light
of frequency $\omega$ propagating along $z$, and excites a collective
chain mode at frequency $\omega_{\mathrm{c}}$.
(b) Normalized (to the geometrical cross section of a single sphere)
absorption cross sections for the 7-NP Ag helix of (a), for RCP (green
curve) and LCP (orange curve) incident light propagating along the helix
axis, in the absence of an excitonic shell. The grey dashed spectrum is
that of the excitonic shell alone, with transition energy $\hbar 
\omega_{\mathrm{exc}} = 2.4$\;eV.
(c) CD spectrum of the uncoupled helix of (a).
(d) Schematic of the helix of (a), now embedded in an excitonic matrix
with a transition at $\omega_{\mathrm{exc}}$, modelled as an ellipsoid
described by the permittivity of eqn~(\ref{eq:eq1}).
(e) Same absorption spectra as in (b), with the excitonic layer present
(with $\hbar \omega_{\mathrm{exc}} = 2.4$\;eV), and
(f) the corresponding CD spectrum.
}\label{fig4}
\end{figure}

We have established so far that excitonic materials can be
successfully employed to reconfigure the chirality response of
a system by means of either amplifying the intensity of the
measured signal, or adding new spectral features, even in the
least sophisticated structures. Among the numerous nanostructures
that have been shown in recent years to demonstrate strong structural
optical chirality, metallic-NP helices, whose optical activity is
encoded in the geometry of the system rather than the optical parameters,
attract considerable interest~\cite{fan_jpcc115,kuzyk_nat483,song_nl13}.
Due to their intense plasmonic near fields, they significantly enhance
chiroptical effects, while also offering unique control and tunability
capabilities provided by DNA-origami assembling~\cite{kuzyk_acsphot5,yesilyurt_aom9},
which has evolved into the main nanofabrication technique
for such systems. Recently, we analyzed how such plasmonic NP helices can
be employed as efficient templates for the evaluation of
quantum-informed models for plasmonics~\cite{tserkezis_acsphot5b}; here
we employ the same architecture as a realistic example of the
reconfigurable chirality discussed above.

In what follows, we consider Ag nanospheres (described by the experimental
permittivity of Johnson and Christy~\cite{johnson_prb6}) of radius
$R = 5$\;nm, a size that lies within the limits of what can realistically
be supported by DNA origami~\cite{liu_chemrev118}. Each helix contains
7 particles, arranged around the $z$ axis, and revolving by steps of
$30^{\circ}$, so that 7 particles constitute one full revolution about
$z$, as shown in the schematics of Fig.~\ref{fig4}(a). The vertical
center-to-center distance of the spheres along $z$ is $2R$. In this
arrangement, the distances between nearest-neighbour NPs are really
small, nearly touching, thus ensuring intense plasmonic interactions,
which are expected to lead to collective plasmonic chain modes at
frequencies $\omega_{\mathrm{c}}$, that propagate along the direction
of the field polarization~\cite{tserkezis_part31}, thus generating
characteristic spectral features that are sensitive to both the NP
arrangement and the incident polarization~\cite{tserkezis_acsphot5b}.
For more details about the simulations, see ESI.

Fig.~\ref{fig4} shows the absorption cross section and CD spectra
for propagation along the helix axis ($z$), as sketched in (a). In panel~(b)
we show the absorption spectra  (normalized to the geometric cross section of an individual NP) in the absence of an excitonic shell for L/RCP polarized
light (orange/green curve), together with the spectrum of the excitonic matrix
alone (grey dashed curve); the matrix is an ellipsoid just enclosing the entire
helix (long axis $30$\;nm, short axis $17.5$\;nm) described by the generic 
Lorentzian of eqn~(\ref{eq:eq1}) with $\varepsilon_{b} = 1$, $f = 0.02$ and
$\hbar \gamma_{\mathrm{exc}} = 0.052$\;eV, while in this particular spectrum
we choose $\hbar \omega_{\mathrm{exc}} = 2.4$\;eV. The absorption spectrum is
characterized by two low-energy resonances, close in energy, labeled as peaks A and B, that correspond
to chain plasmons at energies $\hbar \omega_{\mathrm{c}}$, emerging from the
interaction of individual spheres along the direction of the incident
polarization; since the incident light contains both $x$ and $y$
electric-field components, small sub-chains are excited along both directions,
with slightly smaller energies because of the different effective number of
chains in each direction~\cite{tserkezis_acsphot5b}. Fig.~\ref{fig4} shows a different preference in absorption at the energies of the two resonances; as indicated by the color of the labels, resonance A absorbs RCP light more efficiently than LCP, and the opposite occurs for resonance B. The scattering spectra of
these helices are characterized by similar features, only much weaker in
amplitude, since such small plasmonic NPs are dominated by absorption.
Such modes have an electric and magnetic character simultaneously, since
they emerge from the circulation of the electric dipolar plasmonic modes
of individual NPs, leading thus naturally to strong chiroptical activity.

\begin{figure}[ht]
\centering
\includegraphics[width = 0.8\columnwidth]{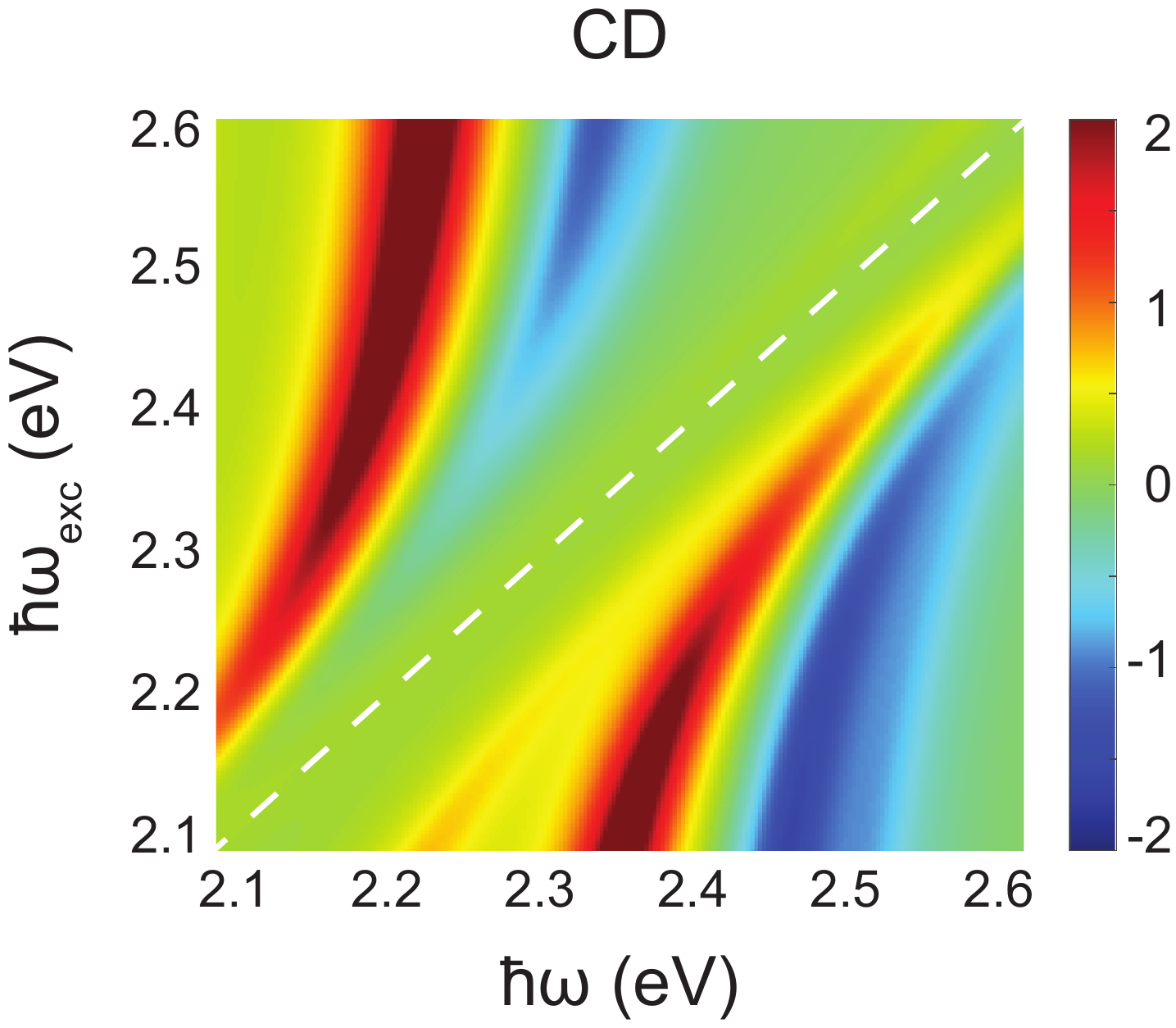}
\caption{
Photon versus exciton energy colour map of the CD corresponding to the
helix of Fig.~\ref{fig4}.
}\label{fig5}
\end{figure}

The resulting CD spectrum is shown in Fig.~\ref{fig4}(c), exhibiting two intense resonant features of opposite signs at around $2.3$\;eV and $2.4$\;eV, with an inflection point between them. These are
exactly the CD features that we will try to manipulate through strong coupling
with the excitonic layer. Fig~\ref{fig4}(d) shows the same helix as in (a),
now embedded in the matrix described above. Depending on its energy, the
excitonic transition of the matrix interacts more strongly with one of the
two helix resonances, leading to its splitting into two peaks. This is shown
for example in Fig.~\ref{fig4}(e), for the excitonic transition depicted in
panel (b); the interaction leads to a wide anticrossing between peaks $\textrm{B}_1$ and $\textrm{B}_2$, of the order of
$300$\;meV. Interestingly, because the second chain resonance, labeled as A in Fig.~\ref{fig4}(b), is also close
in energy, it can also couple to the exciton; at the same time, the excitonic
layer itself supports the geometric resonance C~\cite{antosiewicz_acsphot1} that still manifests as an uncoupled spectral feature at $2.4$\;eV. As a consequence,
the absorption spectrum is now characterized by five resonances, each with
different chirality properties. Tracing which of the split branches corresponds
to which original resonance is straightforward, as the hybrid modes still
maintain the preference in handedness of the uncoupled ones: in this particular example,
the resonances with higher amplitude under RCP incidence ($\textrm{A}_1$, $\textrm{A}_2$) originate from
resonance A, while higher LCP intensity ($\textrm{B}_1$, $\textrm{B}_2$) corresponds to resonance B. Since resonance C is entirely related
to the matrix, it has no handedness and its CD signal is practically zero,
as one can see in Fig.~\ref{fig4}(f).
Comparing Figs.~\ref{fig4}(c) and (f), one can argue that the introduction of the excitonic matrix has led to a reduction of the maximum measured CD signal over the energy window of interest. In principle, this could be intuitively anticipated, since the contribution of an achiral absorbing medium has been added to the total absorption cross sections of 
eqn~(\ref{Eq:CD-DS}a). However, the splitting of the resonant features and the manifestation of new peaks itself, in a way enhances the CD signal, since high values are measured at energies where, otherwise, the uncoupled structure would give zero. Indeed, in Figs.~\ref{fig4}(c) and (f) energy windows of overall enhancement of the chiral response can be identified, e.g. at about $2.2$\;eV, where the valley of nearly zero CD in the uncoupled case [Fig.~\ref{fig4}(c)] becomes a resonance in the case of the coupled system [Fig.~\ref{fig4}(f)] exactly due to the hybridization and the emergence of new absorption resonances and their corresponding features in CD.

By allowing $\omega_{\mathrm{exc}}$ to vary, one can scan the entire energy
window of interest, and obtain the CD colour map of Fig.~\ref{fig5}. Here it
is more evident how the one band of strong CD shown in Fig.~\ref{fig4}(c)
has transformed into two, separated by an anticrossing with practically zero
CD, following the linear dispersion corresponding to the exciton-polariton
resonance of the matrix itself. It is also clear that, due to this hybridization,
one can tune the system so as to exhibit strong optical chirality at energies
up to $0.2$\;eV away from the corresponding window in the uncoupled case,
thus enabling the post-manufacturing manipulation of the chirality of a
structurally chiral object. Finally, it should be noted that similar CD maps
are obtained for other directions of light propagation as well (albeit shifted 
somehow in energy ---longer chains are, in this respect more preferable, since
their response tends to be more isotropic~\cite{tserkezis_acsphot5b}), meaning
that the effect should be visible  even in the case of an ensemble of such
helices immersed in a resonant solution.

\section{Conclusions}
We theoretically explored the possibility of manipulating the chiroptical
response of two distinct nanostructures, with different types of chirality,
through coupling with excitonic materials. The chirality density of a field
is correlated to overlapping parallel electric and magnetic field components.
These components become reconfigurable upon addition of the excitonic material,
which then translates into a reconfigurable signal in the scattering and
absorption properties of the chiral structure, such as DS and CD. The effect
is most evident when the system operates closer to the strong coupling regime,
where the properties of each component are tied together and the coupled
configuration is characterized by hybrid light--matter resonances. We
demonstrated that the introduction of excitonic components can lead to the
amplification of the intensity of the measured signal, or to the emergence
of new spectral features that inherit the handedness of the uncoupled structure.
As a practical example, we focused on plasmonic spheres in a helical arrangement;
nevertheless, our findings can be directly extended to any other system
exhibiting natural, structural, or magnetic chirality.

\section*{Author Contributions}
C.~T., P.~E.~S., and S.~D. conceived the idea.
P.~E.~S. performed the calculations on the single nanoparticle. C.~T.  performed the calculations on the nanoparticle helix.
All authors contributed equally to the discussion of the results and the writing of the paper.

\section*{Conflicts of interest}
There are no conflicts to declare.

\section*{Acknowledgements}

P.~E.~S. is the recipient of the Zonta Denmark's Scholarship for
female PhD students in Science and Technology 2021.
N.~A.~M. is a VILLUM Investigator supported by VILLUM FONDEN
(Grant No.~16498).
G.~P.~A. acknowledges support from the Swiss National Science Foundation (200021\_184687) and National Center of Competence in Research Bio-Inspired Materials NCCR (51NF40\_182881).

\bibliography{rsc}

\begin{thebibliography}{56}%
\makeatletter
\providecommand \@ifxundefined [1]{%
 \@ifx{#1\undefined}
}%
\providecommand \@ifnum [1]{%
 \ifnum #1\expandafter \@firstoftwo
 \else \expandafter \@secondoftwo
 \fi
}%
\providecommand \@ifx [1]{%
 \ifx #1\expandafter \@firstoftwo
 \else \expandafter \@secondoftwo
 \fi
}%
\providecommand \natexlab [1]{#1}%
\providecommand \enquote  [1]{``#1''}%
\providecommand \bibnamefont  [1]{#1}%
\providecommand \bibfnamefont [1]{#1}%
\providecommand \citenamefont [1]{#1}%
\providecommand \href@noop [0]{\@secondoftwo}%
\providecommand \href [0]{\begingroup \@sanitize@url \@href}%
\providecommand \@href[1]{\@@startlink{#1}\@@href}%
\providecommand \@@href[1]{\endgroup#1\@@endlink}%
\providecommand \@sanitize@url [0]{\catcode `\\12\catcode `\$12\catcode
  `\&12\catcode `\#12\catcode `\^12\catcode `\_12\catcode `\%12\relax}%
\providecommand \@@startlink[1]{}%
\providecommand \@@endlink[0]{}%
\providecommand \url  [0]{\begingroup\@sanitize@url \@url }%
\providecommand \@url [1]{\endgroup\@href {#1}{\urlprefix }}%
\providecommand \urlprefix  [0]{URL }%
\providecommand \Eprint [0]{\href }%
\providecommand \doibase [0]{http://dx.doi.org/}%
\providecommand \selectlanguage [0]{\@gobble}%
\providecommand \bibinfo  [0]{\@secondoftwo}%
\providecommand \bibfield  [0]{\@secondoftwo}%
\providecommand \translation [1]{[#1]}%
\providecommand \BibitemOpen [0]{}%
\providecommand \bibitemStop [0]{}%
\providecommand \bibitemNoStop [0]{.\EOS\space}%
\providecommand \EOS [0]{\spacefactor3000\relax}%
\providecommand \BibitemShut  [1]{\csname bibitem#1\endcsname}%
\let\auto@bib@innerbib\@empty
%</preamble>
\bibitem [{\citenamefont {Sun}\ \emph {et~al.}(2018)\citenamefont {Sun},
  \citenamefont {Hao}, \citenamefont {Li}, \citenamefont {Qu}, \citenamefont
  {Xu}, \citenamefont {Hao}, \citenamefont {Xu},\ and\ \citenamefont
  {Kuang}}]{sun_natcom9}%
  \BibitemOpen
  \bibfield  {author} {\bibinfo {author} {\bibfnamefont {M.}~\bibnamefont
  {Sun}}, \bibinfo {author} {\bibfnamefont {T.}~\bibnamefont {Hao}}, \bibinfo
  {author} {\bibfnamefont {X.}~\bibnamefont {Li}}, \bibinfo {author}
  {\bibfnamefont {A.}~\bibnamefont {Qu}}, \bibinfo {author} {\bibfnamefont
  {L.}~\bibnamefont {Xu}}, \bibinfo {author} {\bibfnamefont {C.}~\bibnamefont
  {Hao}}, \bibinfo {author} {\bibfnamefont {C.}~\bibnamefont {Xu}}, \ and\
  \bibinfo {author} {\bibfnamefont {H.}~\bibnamefont {Kuang}},\ }\href
  {\doibase 10.1038/s41467-018-06946-z} {\bibfield  {journal} {\bibinfo
  {journal} {Nat. Commun.}\ }\textbf {\bibinfo {volume} {9}},\ \bibinfo {pages}
  {4494} (\bibinfo {year} {2018})}\BibitemShut {NoStop}%
\bibitem [{\citenamefont {Nguyen}\ \emph {et~al.}(2006)\citenamefont {Nguyen},
  \citenamefont {He},\ and\ \citenamefont {Pham-Huy}}]{nguyen_ijbs2}%
  \BibitemOpen
  \bibfield  {author} {\bibinfo {author} {\bibfnamefont {L.~A.}\ \bibnamefont
  {Nguyen}}, \bibinfo {author} {\bibfnamefont {H.}~\bibnamefont {He}}, \ and\
  \bibinfo {author} {\bibfnamefont {C.}~\bibnamefont {Pham-Huy}},\ }\href@noop
  {} {\bibfield  {journal} {\bibinfo  {journal} {Int. J. Biomed. Sci.}\
  }\textbf {\bibinfo {volume} {2}},\ \bibinfo {pages} {85} (\bibinfo {year}
  {2006})}\BibitemShut {NoStop}%
\bibitem [{\citenamefont {Alvarez-Rivera}\ \emph {et~al.}(2020)\citenamefont
  {Alvarez-Rivera}, \citenamefont {Bueno}, \citenamefont {Ballesteros-Vivas},\
  and\ \citenamefont {Cifuentes}}]{alvarez_trac123}%
  \BibitemOpen
  \bibfield  {author} {\bibinfo {author} {\bibfnamefont {G.}~\bibnamefont
  {Alvarez-Rivera}}, \bibinfo {author} {\bibfnamefont {M.}~\bibnamefont
  {Bueno}}, \bibinfo {author} {\bibfnamefont {D.}~\bibnamefont
  {Ballesteros-Vivas}}, \ and\ \bibinfo {author} {\bibfnamefont
  {A.}~\bibnamefont {Cifuentes}},\ }\href {\doibase 10.1016/j.trac.2019.115761}
  {\bibfield  {journal} {\bibinfo  {journal} {Trends Anal. Chem.}\ }\textbf
  {\bibinfo {volume} {123}},\ \bibinfo {pages} {115761} (\bibinfo {year}
  {2020})}\BibitemShut {NoStop}%
\bibitem [{\citenamefont {Berova}\ \emph {et~al.}(2000)\citenamefont {Berova},
  \citenamefont {Nakanishi},\ and\ \citenamefont {Woody}}]{Berova_Wiley2000}%
  \BibitemOpen
  \bibinfo {editor} {\bibfnamefont {N.}~\bibnamefont {Berova}}, \bibinfo
  {editor} {\bibfnamefont {K.}~\bibnamefont {Nakanishi}}, \ and\ \bibinfo
  {editor} {\bibfnamefont {R.~W.}\ \bibnamefont {Woody}},\ eds.,\ \href@noop {}
  {\emph {\bibinfo {title} {Circular Dichroism: Principles and Applications,
  2nd Ed.}}}\ (\bibinfo  {publisher} {John Wiley and Sons},\ \bibinfo {address}
  {New York},\ \bibinfo {year} {2000})\BibitemShut {NoStop}%
\bibitem [{\citenamefont {Hentschel}\ \emph {et~al.}(2017)\citenamefont
  {Hentschel}, \citenamefont {Sch\"{a}ferling}, \citenamefont {Duan},
  \citenamefont {Giessen},\ and\ \citenamefont {Liu}}]{hentschel_sciadv3}%
  \BibitemOpen
  \bibfield  {author} {\bibinfo {author} {\bibfnamefont {M.}~\bibnamefont
  {Hentschel}}, \bibinfo {author} {\bibfnamefont {M.}~\bibnamefont
  {Sch\"{a}ferling}}, \bibinfo {author} {\bibfnamefont {X.}~\bibnamefont
  {Duan}}, \bibinfo {author} {\bibfnamefont {H.}~\bibnamefont {Giessen}}, \
  and\ \bibinfo {author} {\bibfnamefont {N.}~\bibnamefont {Liu}},\ }\href
  {\doibase 10.1126/sciadv.1602735} {\bibfield  {journal} {\bibinfo  {journal}
  {Sci. Adv.}\ }\textbf {\bibinfo {volume} {3}},\ \bibinfo {pages} {e1602735}
  (\bibinfo {year} {2017})}\BibitemShut {NoStop}%
\bibitem [{\citenamefont {Petronijevic}\ \emph {et~al.}(2022)\citenamefont
  {Petronijevic}, \citenamefont {Belardini}, \citenamefont {Leahu},
  \citenamefont {{Li Voti}},\ and\ \citenamefont
  {Sibilia}}]{petronijevic_omex12}%
  \BibitemOpen
  \bibfield  {author} {\bibinfo {author} {\bibfnamefont {E.}~\bibnamefont
  {Petronijevic}}, \bibinfo {author} {\bibfnamefont {A.}~\bibnamefont
  {Belardini}}, \bibinfo {author} {\bibfnamefont {G.}~\bibnamefont {Leahu}},
  \bibinfo {author} {\bibfnamefont {R.}~\bibnamefont {{Li Voti}}}, \ and\
  \bibinfo {author} {\bibfnamefont {C.}~\bibnamefont {Sibilia}},\ }\href
  {\doibase 10.1364/ome.456496} {\bibfield  {journal} {\bibinfo  {journal}
  {Opt. Mater. Express}\ }\textbf {\bibinfo {volume} {12}},\ \bibinfo {pages}
  {2724} (\bibinfo {year} {2022})}\BibitemShut {NoStop}%
\bibitem [{\citenamefont {Collins}\ \emph {et~al.}(2017)\citenamefont
  {Collins}, \citenamefont {Kuppe}, \citenamefont {Hooper}, \citenamefont
  {Sibilia}, \citenamefont {Centini},\ and\ \citenamefont
  {Valev}}]{collins_aom5}%
  \BibitemOpen
  \bibfield  {author} {\bibinfo {author} {\bibfnamefont {J.~T.}\ \bibnamefont
  {Collins}}, \bibinfo {author} {\bibfnamefont {C.}~\bibnamefont {Kuppe}},
  \bibinfo {author} {\bibfnamefont {D.~C.}\ \bibnamefont {Hooper}}, \bibinfo
  {author} {\bibfnamefont {C.}~\bibnamefont {Sibilia}}, \bibinfo {author}
  {\bibfnamefont {M.}~\bibnamefont {Centini}}, \ and\ \bibinfo {author}
  {\bibfnamefont {V.~K.}\ \bibnamefont {Valev}},\ }\href {\doibase
  10.1002/adom.201700182} {\bibfield  {journal} {\bibinfo  {journal} {Adv. Opt.
  Mater.}\ }\textbf {\bibinfo {volume} {5}},\ \bibinfo {pages} {1700182}
  (\bibinfo {year} {2017})}\BibitemShut {NoStop}%
\bibitem [{\citenamefont {Plum}\ \emph {et~al.}(2009)\citenamefont {Plum},
  \citenamefont {Zhou}, \citenamefont {Dong}, \citenamefont {Fedotov},
  \citenamefont {Koschny}, \citenamefont {Soukoulis},\ and\ \citenamefont
  {Zheludev}}]{plum_prb79}%
  \BibitemOpen
  \bibfield  {author} {\bibinfo {author} {\bibfnamefont {E.}~\bibnamefont
  {Plum}}, \bibinfo {author} {\bibfnamefont {J.}~\bibnamefont {Zhou}}, \bibinfo
  {author} {\bibfnamefont {J.}~\bibnamefont {Dong}}, \bibinfo {author}
  {\bibfnamefont {V.~A.}\ \bibnamefont {Fedotov}}, \bibinfo {author}
  {\bibfnamefont {T.}~\bibnamefont {Koschny}}, \bibinfo {author} {\bibfnamefont
  {C.~M.}\ \bibnamefont {Soukoulis}}, \ and\ \bibinfo {author} {\bibfnamefont
  {N.~I.}\ \bibnamefont {Zheludev}},\ }\href {\doibase
  10.1103/PhysRevB.79.035407} {\bibfield  {journal} {\bibinfo  {journal} {Phys.
  Rev. B}\ }\textbf {\bibinfo {volume} {79}},\ \bibinfo {pages} {035407}
  (\bibinfo {year} {2009})}\BibitemShut {NoStop}%
\bibitem [{\citenamefont {Gansel}\ \emph {et~al.}(2009)\citenamefont {Gansel},
  \citenamefont {Thiel}, \citenamefont {Rill}, \citenamefont {Decker},
  \citenamefont {Bade}, \citenamefont {Saile}, \citenamefont {{von~Freymann}},
  \citenamefont {Linden},\ and\ \citenamefont {Wegener}}]{gansel_sci352}%
  \BibitemOpen
  \bibfield  {author} {\bibinfo {author} {\bibfnamefont {J.~K.}\ \bibnamefont
  {Gansel}}, \bibinfo {author} {\bibfnamefont {M.}~\bibnamefont {Thiel}},
  \bibinfo {author} {\bibfnamefont {M.~S.}\ \bibnamefont {Rill}}, \bibinfo
  {author} {\bibfnamefont {M.}~\bibnamefont {Decker}}, \bibinfo {author}
  {\bibfnamefont {K.}~\bibnamefont {Bade}}, \bibinfo {author} {\bibfnamefont
  {V.}~\bibnamefont {Saile}}, \bibinfo {author} {\bibfnamefont
  {G.}~\bibnamefont {{von~Freymann}}}, \bibinfo {author} {\bibfnamefont
  {S.}~\bibnamefont {Linden}}, \ and\ \bibinfo {author} {\bibfnamefont
  {M.}~\bibnamefont {Wegener}},\ }\href {\doibase 10.1126/science.1177031}
  {\bibfield  {journal} {\bibinfo  {journal} {Science}\ }\textbf {\bibinfo
  {volume} {325}},\ \bibinfo {pages} {1513} (\bibinfo {year}
  {2009})}\BibitemShut {NoStop}%
\bibitem [{\citenamefont {Lodahl}\ \emph {et~al.}(2017)\citenamefont {Lodahl},
  \citenamefont {Mahmoodian}, \citenamefont {Stobbe}, \citenamefont
  {Rauschenbeutel}, \citenamefont {Schneeweiss}, \citenamefont {Volz},
  \citenamefont {Pichler},\ and\ \citenamefont {Zoller}}]{lodahl_nat541}%
  \BibitemOpen
  \bibfield  {author} {\bibinfo {author} {\bibfnamefont {P.}~\bibnamefont
  {Lodahl}}, \bibinfo {author} {\bibfnamefont {S.}~\bibnamefont {Mahmoodian}},
  \bibinfo {author} {\bibfnamefont {S.}~\bibnamefont {Stobbe}}, \bibinfo
  {author} {\bibfnamefont {A.}~\bibnamefont {Rauschenbeutel}}, \bibinfo
  {author} {\bibfnamefont {P.}~\bibnamefont {Schneeweiss}}, \bibinfo {author}
  {\bibfnamefont {J.}~\bibnamefont {Volz}}, \bibinfo {author} {\bibfnamefont
  {H.}~\bibnamefont {Pichler}}, \ and\ \bibinfo {author} {\bibfnamefont
  {P.}~\bibnamefont {Zoller}},\ }\href {\doibase 10.1038/nature21037}
  {\bibfield  {journal} {\bibinfo  {journal} {Nature}\ }\textbf {\bibinfo
  {volume} {541}},\ \bibinfo {pages} {473} (\bibinfo {year}
  {2017})}\BibitemShut {NoStop}%
\bibitem [{\citenamefont {Aiello}\ \emph {et~al.}(2022)\citenamefont {Aiello},
  \citenamefont {Abendroth}, \citenamefont {Abbas}, \citenamefont {Afanasev},
  \citenamefont {Agarwal}, \citenamefont {Banerjee}, \citenamefont {Beratan},
  \citenamefont {Belling}, \citenamefont {Berche}, \citenamefont {Botana},
  \citenamefont {Caram}, \citenamefont {Celardo}, \citenamefont {Cuniberti},
  \citenamefont {Garcia-Etxarri}, \citenamefont {Dianat}, \citenamefont
  {Diez-Perez}, \citenamefont {Guo}, \citenamefont {Gutierrez}, \citenamefont
  {Herrmann}, \citenamefont {Hihath}, \citenamefont {Kale}, \citenamefont
  {Kurian}, \citenamefont {Lai}, \citenamefont {Liu}, \citenamefont {Lopez},
  \citenamefont {Medina}, \citenamefont {Mujica}, \citenamefont {Naaman},
  \citenamefont {Noormandipour}, \citenamefont {Palma}, \citenamefont
  {Paltiel}, \citenamefont {Petuskey}, \citenamefont {Ribeiro-Silva},
  \citenamefont {Saenz}, \citenamefont {Santos}, \citenamefont
  {Solyanik-Gorgone}, \citenamefont {Sorger}, \citenamefont {Stemer},
  \citenamefont {Ugalde}, \citenamefont {Valdes-Curiel}, \citenamefont
  {Varela}, \citenamefont {Waldeck}, \citenamefont {Wasielewski}, \citenamefont
  {Weiss}, \citenamefont {Zacharias},\ and\ \citenamefont
  {Wang}}]{aiello_nn16}%
  \BibitemOpen
  \bibfield  {author} {\bibinfo {author} {\bibfnamefont {C.~D.}\ \bibnamefont
  {Aiello}}, \bibinfo {author} {\bibfnamefont {J.~M.}\ \bibnamefont
  {Abendroth}}, \bibinfo {author} {\bibfnamefont {M.}~\bibnamefont {Abbas}},
  \bibinfo {author} {\bibfnamefont {A.}~\bibnamefont {Afanasev}}, \bibinfo
  {author} {\bibfnamefont {S.}~\bibnamefont {Agarwal}}, \bibinfo {author}
  {\bibfnamefont {A.~S.}\ \bibnamefont {Banerjee}}, \bibinfo {author}
  {\bibfnamefont {D.~N.}\ \bibnamefont {Beratan}}, \bibinfo {author}
  {\bibfnamefont {J.~N.}\ \bibnamefont {Belling}}, \bibinfo {author}
  {\bibfnamefont {B.}~\bibnamefont {Berche}}, \bibinfo {author} {\bibfnamefont
  {A.}~\bibnamefont {Botana}}, \bibinfo {author} {\bibfnamefont {J.~R.}\
  \bibnamefont {Caram}}, \bibinfo {author} {\bibfnamefont {G.~L.}\ \bibnamefont
  {Celardo}}, \bibinfo {author} {\bibfnamefont {G.}~\bibnamefont {Cuniberti}},
  \bibinfo {author} {\bibfnamefont {A.}~\bibnamefont {Garcia-Etxarri}},
  \bibinfo {author} {\bibfnamefont {A.}~\bibnamefont {Dianat}}, \bibinfo
  {author} {\bibfnamefont {I.}~\bibnamefont {Diez-Perez}}, \bibinfo {author}
  {\bibfnamefont {Y.}~\bibnamefont {Guo}}, \bibinfo {author} {\bibfnamefont
  {R.}~\bibnamefont {Gutierrez}}, \bibinfo {author} {\bibfnamefont
  {C.}~\bibnamefont {Herrmann}}, \bibinfo {author} {\bibfnamefont
  {J.}~\bibnamefont {Hihath}}, \bibinfo {author} {\bibfnamefont
  {S.}~\bibnamefont {Kale}}, \bibinfo {author} {\bibfnamefont {P.}~\bibnamefont
  {Kurian}}, \bibinfo {author} {\bibfnamefont {Y.-C.}\ \bibnamefont {Lai}},
  \bibinfo {author} {\bibfnamefont {T.}~\bibnamefont {Liu}}, \bibinfo {author}
  {\bibfnamefont {A.}~\bibnamefont {Lopez}}, \bibinfo {author} {\bibfnamefont
  {E.}~\bibnamefont {Medina}}, \bibinfo {author} {\bibfnamefont
  {V.}~\bibnamefont {Mujica}}, \bibinfo {author} {\bibfnamefont
  {R.}~\bibnamefont {Naaman}}, \bibinfo {author} {\bibfnamefont
  {M.}~\bibnamefont {Noormandipour}}, \bibinfo {author} {\bibfnamefont {J.~L.}\
  \bibnamefont {Palma}}, \bibinfo {author} {\bibfnamefont {Y.}~\bibnamefont
  {Paltiel}}, \bibinfo {author} {\bibfnamefont {W.}~\bibnamefont {Petuskey}},
  \bibinfo {author} {\bibfnamefont {J.~C.}\ \bibnamefont {Ribeiro-Silva}},
  \bibinfo {author} {\bibfnamefont {J.~J.}\ \bibnamefont {Saenz}}, \bibinfo
  {author} {\bibfnamefont {E.~J.~G.}\ \bibnamefont {Santos}}, \bibinfo {author}
  {\bibfnamefont {M.}~\bibnamefont {Solyanik-Gorgone}}, \bibinfo {author}
  {\bibfnamefont {V.~J.}\ \bibnamefont {Sorger}}, \bibinfo {author}
  {\bibfnamefont {D.~M.}\ \bibnamefont {Stemer}}, \bibinfo {author}
  {\bibfnamefont {J.~M.}\ \bibnamefont {Ugalde}}, \bibinfo {author}
  {\bibfnamefont {A.}~\bibnamefont {Valdes-Curiel}}, \bibinfo {author}
  {\bibfnamefont {S.}~\bibnamefont {Varela}}, \bibinfo {author} {\bibfnamefont
  {D.~H.}\ \bibnamefont {Waldeck}}, \bibinfo {author} {\bibfnamefont {M.~R.}\
  \bibnamefont {Wasielewski}}, \bibinfo {author} {\bibfnamefont {P.~S.}\
  \bibnamefont {Weiss}}, \bibinfo {author} {\bibfnamefont {H.}~\bibnamefont
  {Zacharias}}, \ and\ \bibinfo {author} {\bibfnamefont {Q.~H.}\ \bibnamefont
  {Wang}},\ }\href {\doibase 10.1021/acsnano.1c01347} {\bibfield  {journal}
  {\bibinfo  {journal} {ACS Nano}\ }\textbf {\bibinfo {volume} {16}},\ \bibinfo
  {pages} {4989} (\bibinfo {year} {2022})}\BibitemShut {NoStop}%
\bibitem [{\citenamefont {Dolamic}\ \emph {et~al.}(2012)\citenamefont
  {Dolamic}, \citenamefont {Knoppe}, \citenamefont {Dass},\ and\ \citenamefont
  {B{\"u}rgi}}]{dolamic_natcom3}%
  \BibitemOpen
  \bibfield  {author} {\bibinfo {author} {\bibfnamefont {I.}~\bibnamefont
  {Dolamic}}, \bibinfo {author} {\bibfnamefont {S.}~\bibnamefont {Knoppe}},
  \bibinfo {author} {\bibfnamefont {A.}~\bibnamefont {Dass}}, \ and\ \bibinfo
  {author} {\bibfnamefont {T.}~\bibnamefont {B{\"u}rgi}},\ }\href {\doibase
  10.1038/ncomms180} {\bibfield  {journal} {\bibinfo  {journal} {Nat. Commun.}\
  }\textbf {\bibinfo {volume} {3}},\ \bibinfo {pages} {798} (\bibinfo {year}
  {2012})}\BibitemShut {NoStop}%
\bibitem [{\citenamefont {Lan}\ \emph {et~al.}(2015)\citenamefont {Lan},
  \citenamefont {Lu}, \citenamefont {Shen}, \citenamefont {Ke}, \citenamefont
  {Ni},\ and\ \citenamefont {Wang}}]{lan_jacs137}%
  \BibitemOpen
  \bibfield  {author} {\bibinfo {author} {\bibfnamefont {X.}~\bibnamefont
  {Lan}}, \bibinfo {author} {\bibfnamefont {X.}~\bibnamefont {Lu}}, \bibinfo
  {author} {\bibfnamefont {C.}~\bibnamefont {Shen}}, \bibinfo {author}
  {\bibfnamefont {Y.}~\bibnamefont {Ke}}, \bibinfo {author} {\bibfnamefont
  {W.}~\bibnamefont {Ni}}, \ and\ \bibinfo {author} {\bibfnamefont
  {Q.}~\bibnamefont {Wang}},\ }\href {\doibase 10.1021/ja511333q} {\bibfield
  {journal} {\bibinfo  {journal} {J. Am. Chem. Soc.}\ }\textbf {\bibinfo
  {volume} {137}},\ \bibinfo {pages} {457} (\bibinfo {year}
  {2015})}\BibitemShut {NoStop}%
\bibitem [{\citenamefont {Droulias}(2020)}]{droulias_prb102}%
  \BibitemOpen
  \bibfield  {author} {\bibinfo {author} {\bibfnamefont {S.}~\bibnamefont
  {Droulias}},\ }\href {\doibase 10.1103/PhysRevB.102.075119} {\bibfield
  {journal} {\bibinfo  {journal} {Phys. Rev. B}\ }\textbf {\bibinfo {volume}
  {102}},\ \bibinfo {pages} {075119} (\bibinfo {year} {2020})}\BibitemShut
  {NoStop}%
\bibitem [{\citenamefont {Maccaferri}\ \emph {et~al.}(2020)\citenamefont
  {Maccaferri}, \citenamefont {Zubritskaya}, \citenamefont {Razdolski},
  \citenamefont {Chioar}, \citenamefont {Belotelov}, \citenamefont {Kapaklis},
  \citenamefont {Oppeneer},\ and\ \citenamefont
  {Dmitriev}}]{maccaferri_jap127}%
  \BibitemOpen
  \bibfield  {author} {\bibinfo {author} {\bibfnamefont {N.}~\bibnamefont
  {Maccaferri}}, \bibinfo {author} {\bibfnamefont {I.}~\bibnamefont
  {Zubritskaya}}, \bibinfo {author} {\bibfnamefont {I.}~\bibnamefont
  {Razdolski}}, \bibinfo {author} {\bibfnamefont {I.-A.}\ \bibnamefont
  {Chioar}}, \bibinfo {author} {\bibfnamefont {V.}~\bibnamefont {Belotelov}},
  \bibinfo {author} {\bibfnamefont {V.}~\bibnamefont {Kapaklis}}, \bibinfo
  {author} {\bibfnamefont {P.~M.}\ \bibnamefont {Oppeneer}}, \ and\ \bibinfo
  {author} {\bibfnamefont {A.}~\bibnamefont {Dmitriev}},\ }\href {\doibase
  10.1063/1.5100826} {\bibfield  {journal} {\bibinfo  {journal} {J. Appl.
  Phys.}\ }\textbf {\bibinfo {volume} {127}},\ \bibinfo {pages} {080903}
  (\bibinfo {year} {2020})}\BibitemShut {NoStop}%
\bibitem [{\citenamefont {Caridad}\ \emph {et~al.}(2021)\citenamefont
  {Caridad}, \citenamefont {Tserkezis}, \citenamefont {Santos}, \citenamefont
  {Plochocka}, \citenamefont {Venkatesan}, \citenamefont {Coey}, \citenamefont
  {Mortensen}, \citenamefont {Rikken},\ and\ \citenamefont
  {Krsti\'{c}}}]{caridad_prl126}%
  \BibitemOpen
  \bibfield  {author} {\bibinfo {author} {\bibfnamefont {J.~M.}\ \bibnamefont
  {Caridad}}, \bibinfo {author} {\bibfnamefont {C.}~\bibnamefont {Tserkezis}},
  \bibinfo {author} {\bibfnamefont {J.~E.}\ \bibnamefont {Santos}}, \bibinfo
  {author} {\bibfnamefont {P.}~\bibnamefont {Plochocka}}, \bibinfo {author}
  {\bibfnamefont {M.}~\bibnamefont {Venkatesan}}, \bibinfo {author}
  {\bibfnamefont {J.~M.~D.}\ \bibnamefont {Coey}}, \bibinfo {author}
  {\bibfnamefont {N.~A.}\ \bibnamefont {Mortensen}}, \bibinfo {author}
  {\bibfnamefont {G.~L.~J.~A.}\ \bibnamefont {Rikken}}, \ and\ \bibinfo
  {author} {\bibfnamefont {V.}~\bibnamefont {Krsti\'{c}}},\ }\href {\doibase
  10.1103/PhysRevLett.126.177401} {\bibfield  {journal} {\bibinfo  {journal}
  {Phys. Rev. Lett.}\ }\textbf {\bibinfo {volume} {126}},\ \bibinfo {pages}
  {177401} (\bibinfo {year} {2021})}\BibitemShut {NoStop}%
\bibitem [{\citenamefont {Stamatopoulou}\ \emph {et~al.}(2020)\citenamefont
  {Stamatopoulou}, \citenamefont {Yannopapas}, \citenamefont {Mortensen},\ and\
  \citenamefont {Tserkezis}}]{stamatopoulou_prb102}%
  \BibitemOpen
  \bibfield  {author} {\bibinfo {author} {\bibfnamefont {P.~E.}\ \bibnamefont
  {Stamatopoulou}}, \bibinfo {author} {\bibfnamefont {V.}~\bibnamefont
  {Yannopapas}}, \bibinfo {author} {\bibfnamefont {N.~A.}\ \bibnamefont
  {Mortensen}}, \ and\ \bibinfo {author} {\bibfnamefont {C.}~\bibnamefont
  {Tserkezis}},\ }\href {\doibase 10.1103/PhysRevB.102.195415} {\bibfield
  {journal} {\bibinfo  {journal} {Phys. Rev. B}\ }\textbf {\bibinfo {volume}
  {102}},\ \bibinfo {pages} {195415} (\bibinfo {year} {2020})}\BibitemShut
  {NoStop}%
\bibitem [{\citenamefont {Barron}(2004)}]{Barron_Cambridge2004}%
  \BibitemOpen
  \bibfield  {author} {\bibinfo {author} {\bibfnamefont {L.~D.}\ \bibnamefont
  {Barron}},\ }\href@noop {} {\emph {\bibinfo {title} {Molecular Light
  Scattering and Optical Activity}}}\ (\bibinfo  {publisher} {Cambridge
  University Press},\ \bibinfo {address} {Cambridge},\ \bibinfo {year}
  {2004})\BibitemShut {NoStop}%
\bibitem [{\citenamefont {Warning}\ \emph {et~al.}(2021)\citenamefont
  {Warning}, \citenamefont {Miandashti}, \citenamefont {{McCarthy}},
  \citenamefont {Zhang}, \citenamefont {Landes},\ and\ \citenamefont
  {Link}}]{warning_nn15}%
  \BibitemOpen
  \bibfield  {author} {\bibinfo {author} {\bibfnamefont {L.~A.}\ \bibnamefont
  {Warning}}, \bibinfo {author} {\bibfnamefont {A.~R.}\ \bibnamefont
  {Miandashti}}, \bibinfo {author} {\bibfnamefont {L.~A.}\ \bibnamefont
  {{McCarthy}}}, \bibinfo {author} {\bibfnamefont {Q.}~\bibnamefont {Zhang}},
  \bibinfo {author} {\bibfnamefont {C.~F.}\ \bibnamefont {Landes}}, \ and\
  \bibinfo {author} {\bibfnamefont {S.}~\bibnamefont {Link}},\ }\href {\doibase
  10.1021/acsnano.1c04992} {\bibfield  {journal} {\bibinfo  {journal} {ACS
  Nano}\ }\textbf {\bibinfo {volume} {15}},\ \bibinfo {pages} {15538} (\bibinfo
  {year} {2021})}\BibitemShut {NoStop}%
\bibitem [{\citenamefont {Hendry}\ \emph {et~al.}(2010)\citenamefont {Hendry},
  \citenamefont {Carpy}, \citenamefont {Johnston}, \citenamefont {Popland},
  \citenamefont {Mikhaylovskiy}, \citenamefont {Lapthorn}, \citenamefont
  {Kelly}, \citenamefont {Barron}, \citenamefont {Gadegaard},\ and\
  \citenamefont {Kadodwala}}]{hendry_natnano5}%
  \BibitemOpen
  \bibfield  {author} {\bibinfo {author} {\bibfnamefont {E.}~\bibnamefont
  {Hendry}}, \bibinfo {author} {\bibfnamefont {T.}~\bibnamefont {Carpy}},
  \bibinfo {author} {\bibfnamefont {J.}~\bibnamefont {Johnston}}, \bibinfo
  {author} {\bibfnamefont {M.}~\bibnamefont {Popland}}, \bibinfo {author}
  {\bibfnamefont {R.~V.}\ \bibnamefont {Mikhaylovskiy}}, \bibinfo {author}
  {\bibfnamefont {A.~J.}\ \bibnamefont {Lapthorn}}, \bibinfo {author}
  {\bibfnamefont {S.~M.}\ \bibnamefont {Kelly}}, \bibinfo {author}
  {\bibfnamefont {L.~D.}\ \bibnamefont {Barron}}, \bibinfo {author}
  {\bibfnamefont {N.}~\bibnamefont {Gadegaard}}, \ and\ \bibinfo {author}
  {\bibfnamefont {M.}~\bibnamefont {Kadodwala}},\ }\href {\doibase
  10.1038/nnao.2010.209} {\bibfield  {journal} {\bibinfo  {journal} {Nat.
  Nanotechnol.}\ }\textbf {\bibinfo {volume} {5}},\ \bibinfo {pages} {783}
  (\bibinfo {year} {2010})}\BibitemShut {NoStop}%
\bibitem [{\citenamefont {Droulias}\ and\ \citenamefont
  {Bougas}(2020)}]{droulias_nl20}%
  \BibitemOpen
  \bibfield  {author} {\bibinfo {author} {\bibfnamefont {S.}~\bibnamefont
  {Droulias}}\ and\ \bibinfo {author} {\bibfnamefont {L.}~\bibnamefont
  {Bougas}},\ }\href {\doibase 10.1021/acs.nanolett.0c01938} {\bibfield
  {journal} {\bibinfo  {journal} {Nano Lett.}\ }\textbf {\bibinfo {volume}
  {20}},\ \bibinfo {pages} {5960} (\bibinfo {year} {2020})}\BibitemShut
  {NoStop}%
\bibitem [{\citenamefont {Valev}\ \emph {et~al.}(2013)\citenamefont {Valev},
  \citenamefont {Baumberg}, \citenamefont {Sibilia},\ and\ \citenamefont
  {Verbiest}}]{valev_admat25}%
  \BibitemOpen
  \bibfield  {author} {\bibinfo {author} {\bibfnamefont {V.~K.}\ \bibnamefont
  {Valev}}, \bibinfo {author} {\bibfnamefont {J.~J.}\ \bibnamefont {Baumberg}},
  \bibinfo {author} {\bibfnamefont {C.}~\bibnamefont {Sibilia}}, \ and\
  \bibinfo {author} {\bibfnamefont {T.}~\bibnamefont {Verbiest}},\ }\href
  {\doibase 10.1002/adma.201205178} {\bibfield  {journal} {\bibinfo  {journal}
  {Adv. Mater.}\ }\textbf {\bibinfo {volume} {25}},\ \bibinfo {pages} {2517}
  (\bibinfo {year} {2013})}\BibitemShut {NoStop}%
\bibitem [{\citenamefont {Mu}\ \emph {et~al.}(2021)\citenamefont {Mu},
  \citenamefont {Hu}, \citenamefont {Cheng}, \citenamefont {Fang},\ and\
  \citenamefont {Sun}}]{mu_ns13}%
  \BibitemOpen
  \bibfield  {author} {\bibinfo {author} {\bibfnamefont {X.}~\bibnamefont
  {Mu}}, \bibinfo {author} {\bibfnamefont {L.}~\bibnamefont {Hu}}, \bibinfo
  {author} {\bibfnamefont {Y.}~\bibnamefont {Cheng}}, \bibinfo {author}
  {\bibfnamefont {Y.}~\bibnamefont {Fang}}, \ and\ \bibinfo {author}
  {\bibfnamefont {M.}~\bibnamefont {Sun}},\ }\href {\doibase
  10.1039/D0NR06272C} {\bibfield  {journal} {\bibinfo  {journal} {Nanoscale}\
  }\textbf {\bibinfo {volume} {13}},\ \bibinfo {pages} {581} (\bibinfo {year}
  {2021})}\BibitemShut {NoStop}%
\bibitem [{\citenamefont {Garc\'{i}a-Etxarri}\ and\ \citenamefont
  {Dionne}(2013)}]{etxarri_prb87}%
  \BibitemOpen
  \bibfield  {author} {\bibinfo {author} {\bibfnamefont {A.}~\bibnamefont
  {Garc\'{i}a-Etxarri}}\ and\ \bibinfo {author} {\bibfnamefont {J.~A.}\
  \bibnamefont {Dionne}},\ }\href {\doibase 10.1103/PhysRevB.87.235409}
  {\bibfield  {journal} {\bibinfo  {journal} {Phys. Rev. B}\ }\textbf {\bibinfo
  {volume} {87}},\ \bibinfo {pages} {235409} (\bibinfo {year}
  {2013})}\BibitemShut {NoStop}%
\bibitem [{\citenamefont {Lasa-Alonso}\ \emph {et~al.}(2020)\citenamefont
  {Lasa-Alonso}, \citenamefont {Abujetas}, \citenamefont {Nodar}, \citenamefont
  {Dionne}, \citenamefont {S{\'a}enz}, \citenamefont {Molina-Terriza},
  \citenamefont {Aizpurua},\ and\ \citenamefont
  {Garc{\'\i}a-Etxarri}}]{lasa_acsp7}%
  \BibitemOpen
  \bibfield  {author} {\bibinfo {author} {\bibfnamefont {J.}~\bibnamefont
  {Lasa-Alonso}}, \bibinfo {author} {\bibfnamefont {D.~R.}\ \bibnamefont
  {Abujetas}}, \bibinfo {author} {\bibfnamefont {{\'A}.}~\bibnamefont {Nodar}},
  \bibinfo {author} {\bibfnamefont {J.~A.}\ \bibnamefont {Dionne}}, \bibinfo
  {author} {\bibfnamefont {J.~J.}\ \bibnamefont {S{\'a}enz}}, \bibinfo {author}
  {\bibfnamefont {G.}~\bibnamefont {Molina-Terriza}}, \bibinfo {author}
  {\bibfnamefont {J.}~\bibnamefont {Aizpurua}}, \ and\ \bibinfo {author}
  {\bibfnamefont {A.}~\bibnamefont {Garc{\'\i}a-Etxarri}},\ }\href {\doibase
  10.1021/acsphotonics.0c00611} {\bibfield  {journal} {\bibinfo  {journal} {ACS
  Photonics}\ }\textbf {\bibinfo {volume} {7}},\ \bibinfo {pages} {2978}
  (\bibinfo {year} {2020})}\BibitemShut {NoStop}%
\bibitem [{\citenamefont {Graf}\ \emph {et~al.}(2019)\citenamefont {Graf},
  \citenamefont {Feis}, \citenamefont {Garcia-Santiago}, \citenamefont
  {Wegener}, \citenamefont {Rockstuhl},\ and\ \citenamefont
  {Fernandez-Corbaton}}]{graf_acsp6}%
  \BibitemOpen
  \bibfield  {author} {\bibinfo {author} {\bibfnamefont {F.}~\bibnamefont
  {Graf}}, \bibinfo {author} {\bibfnamefont {J.}~\bibnamefont {Feis}}, \bibinfo
  {author} {\bibfnamefont {X.}~\bibnamefont {Garcia-Santiago}}, \bibinfo
  {author} {\bibfnamefont {M.}~\bibnamefont {Wegener}}, \bibinfo {author}
  {\bibfnamefont {C.}~\bibnamefont {Rockstuhl}}, \ and\ \bibinfo {author}
  {\bibfnamefont {I.}~\bibnamefont {Fernandez-Corbaton}},\ }\href {\doibase
  10.1021/acsphotonics.8b01454} {\bibfield  {journal} {\bibinfo  {journal} {ACS
  Photonics}\ }\textbf {\bibinfo {volume} {6}},\ \bibinfo {pages} {482}
  (\bibinfo {year} {2019})}\BibitemShut {NoStop}%
\bibitem [{\citenamefont {Kuzyk}\ \emph {et~al.}(2012)\citenamefont {Kuzyk},
  \citenamefont {Schreiber}, \citenamefont {Fan}, \citenamefont {Pardatscher},
  \citenamefont {Roller}, \citenamefont {H\"{o}gele}, \citenamefont {Simmel},
  \citenamefont {Govorov},\ and\ \citenamefont {Liedl}}]{kuzyk_nat483}%
  \BibitemOpen
  \bibfield  {author} {\bibinfo {author} {\bibfnamefont {A.}~\bibnamefont
  {Kuzyk}}, \bibinfo {author} {\bibfnamefont {R.}~\bibnamefont {Schreiber}},
  \bibinfo {author} {\bibfnamefont {Z.}~\bibnamefont {Fan}}, \bibinfo {author}
  {\bibfnamefont {G.}~\bibnamefont {Pardatscher}}, \bibinfo {author}
  {\bibfnamefont {E.-M.}\ \bibnamefont {Roller}}, \bibinfo {author}
  {\bibfnamefont {A.}~\bibnamefont {H\"{o}gele}}, \bibinfo {author}
  {\bibfnamefont {F.~C.}\ \bibnamefont {Simmel}}, \bibinfo {author}
  {\bibfnamefont {A.~O.}\ \bibnamefont {Govorov}}, \ and\ \bibinfo {author}
  {\bibfnamefont {T.}~\bibnamefont {Liedl}},\ }\href {\doibase
  10.1038/nature10889} {\bibfield  {journal} {\bibinfo  {journal} {Nature}\
  }\textbf {\bibinfo {volume} {483}},\ \bibinfo {pages} {311} (\bibinfo {year}
  {2012})}\BibitemShut {NoStop}%
\bibitem [{\citenamefont {Govorov}(2011)}]{govorov_jpcc115}%
  \BibitemOpen
  \bibfield  {author} {\bibinfo {author} {\bibfnamefont {A.~O.}\ \bibnamefont
  {Govorov}},\ }\href {\doibase 10.1021/jp1121432} {\bibfield  {journal}
  {\bibinfo  {journal} {J. Phys. Chem. C}\ }\textbf {\bibinfo {volume} {115}},\
  \bibinfo {pages} {7914} (\bibinfo {year} {2011})}\BibitemShut {NoStop}%
\bibitem [{\citenamefont {Maoz}\ \emph {et~al.}(2012)\citenamefont {Maoz},
  \citenamefont {van~der Weegen}, \citenamefont {Fan}, \citenamefont {Govorov},
  \citenamefont {Ellestad}, \citenamefont {Berova}, \citenamefont {Meijer},\
  and\ \citenamefont {Markovich}}]{maoz_jacs134}%
  \BibitemOpen
  \bibfield  {author} {\bibinfo {author} {\bibfnamefont {B.~M.}\ \bibnamefont
  {Maoz}}, \bibinfo {author} {\bibfnamefont {R.}~\bibnamefont {van~der
  Weegen}}, \bibinfo {author} {\bibfnamefont {Z.}~\bibnamefont {Fan}}, \bibinfo
  {author} {\bibfnamefont {A.~O.}\ \bibnamefont {Govorov}}, \bibinfo {author}
  {\bibfnamefont {G.}~\bibnamefont {Ellestad}}, \bibinfo {author}
  {\bibfnamefont {N.}~\bibnamefont {Berova}}, \bibinfo {author} {\bibfnamefont
  {E.}~\bibnamefont {Meijer}}, \ and\ \bibinfo {author} {\bibfnamefont
  {G.}~\bibnamefont {Markovich}},\ }\href {\doibase 10.1021/ja309016k}
  {\bibfield  {journal} {\bibinfo  {journal} {J. Am. Chem. Soc.}\ }\textbf
  {\bibinfo {volume} {134}},\ \bibinfo {pages} {17807} (\bibinfo {year}
  {2012})}\BibitemShut {NoStop}%
\bibitem [{\citenamefont {Duan}\ \emph {et~al.}(2015)\citenamefont {Duan},
  \citenamefont {Yue},\ and\ \citenamefont {Liu}}]{duan_ns7}%
  \BibitemOpen
  \bibfield  {author} {\bibinfo {author} {\bibfnamefont {X.}~\bibnamefont
  {Duan}}, \bibinfo {author} {\bibfnamefont {S.}~\bibnamefont {Yue}}, \ and\
  \bibinfo {author} {\bibfnamefont {N.}~\bibnamefont {Liu}},\ }\href {\doibase
  10.1039/C5NR04050G} {\bibfield  {journal} {\bibinfo  {journal} {Nanoscale}\
  }\textbf {\bibinfo {volume} {7}},\ \bibinfo {pages} {17237} (\bibinfo {year}
  {2015})}\BibitemShut {NoStop}%
\bibitem [{\citenamefont {Solomon}\ \emph {et~al.}(2018)\citenamefont
  {Solomon}, \citenamefont {Hu}, \citenamefont {Lawrence}, \citenamefont
  {Garc\'{i}a-Etxarri},\ and\ \citenamefont {Dionne}}]{solomon_acsphot6}%
  \BibitemOpen
  \bibfield  {author} {\bibinfo {author} {\bibfnamefont {M.~L.}\ \bibnamefont
  {Solomon}}, \bibinfo {author} {\bibfnamefont {J.}~\bibnamefont {Hu}},
  \bibinfo {author} {\bibfnamefont {M.}~\bibnamefont {Lawrence}}, \bibinfo
  {author} {\bibfnamefont {A.}~\bibnamefont {Garc\'{i}a-Etxarri}}, \ and\
  \bibinfo {author} {\bibfnamefont {J.~A.}\ \bibnamefont {Dionne}},\ }\href
  {\doibase 10.1021/acsphotonics.8b01365} {\bibfield  {journal} {\bibinfo
  {journal} {ACS Photonics}\ }\textbf {\bibinfo {volume} {6}},\ \bibinfo
  {pages} {43} (\bibinfo {year} {2018})}\BibitemShut {NoStop}%
\bibitem [{\citenamefont {Raziman}\ \emph {et~al.}(2019)\citenamefont
  {Raziman}, \citenamefont {Godiksen}, \citenamefont {M\"{u}ller},\ and\
  \citenamefont {Curto}}]{raziman_acsphot6}%
  \BibitemOpen
  \bibfield  {author} {\bibinfo {author} {\bibfnamefont {T.~V.}\ \bibnamefont
  {Raziman}}, \bibinfo {author} {\bibfnamefont {R.~H.}\ \bibnamefont
  {Godiksen}}, \bibinfo {author} {\bibfnamefont {M.~A.}\ \bibnamefont
  {M\"{u}ller}}, \ and\ \bibinfo {author} {\bibfnamefont {A.~G.}\ \bibnamefont
  {Curto}},\ }\href {\doibase 10.1021/acsphotonics.9b01200} {\bibfield
  {journal} {\bibinfo  {journal} {ACS Photonics}\ }\textbf {\bibinfo {volume}
  {6}},\ \bibinfo {pages} {2583} (\bibinfo {year} {2019})}\BibitemShut
  {NoStop}%
\bibitem [{\citenamefont {Zhang}\ \emph {et~al.}(2017)\citenamefont {Zhang},
  \citenamefont {Wu}, \citenamefont {Wang},\ and\ \citenamefont
  {Zhang}}]{zhang_ns17}%
  \BibitemOpen
  \bibfield  {author} {\bibinfo {author} {\bibfnamefont {W.}~\bibnamefont
  {Zhang}}, \bibinfo {author} {\bibfnamefont {T.}~\bibnamefont {Wu}}, \bibinfo
  {author} {\bibfnamefont {R.}~\bibnamefont {Wang}}, \ and\ \bibinfo {author}
  {\bibfnamefont {X.}~\bibnamefont {Zhang}},\ }\href {\doibase
  10.1039/C7NR01527E} {\bibfield  {journal} {\bibinfo  {journal} {Nanoscale}\
  }\textbf {\bibinfo {volume} {9}},\ \bibinfo {pages} {5701} (\bibinfo {year}
  {2017})}\BibitemShut {NoStop}%
\bibitem [{\citenamefont {Yao}\ and\ \citenamefont {Liu}(2018)}]{yao_ns10}%
  \BibitemOpen
  \bibfield  {author} {\bibinfo {author} {\bibfnamefont {K.}~\bibnamefont
  {Yao}}\ and\ \bibinfo {author} {\bibfnamefont {Y.}~\bibnamefont {Liu}},\
  }\href {\doibase 10.1039/C8NR00902C} {\bibfield  {journal} {\bibinfo
  {journal} {Nanoscale}\ }\textbf {\bibinfo {volume} {10}},\ \bibinfo {pages}
  {8779} (\bibinfo {year} {2018})}\BibitemShut {NoStop}%
\bibitem [{\citenamefont {Fofang}\ \emph {et~al.}(2008)\citenamefont {Fofang},
  \citenamefont {Park}, \citenamefont {Neumann}, \citenamefont {Mirin},
  \citenamefont {Nordlander},\ and\ \citenamefont {Halas}}]{fofang_nl10}%
  \BibitemOpen
  \bibfield  {author} {\bibinfo {author} {\bibfnamefont {N.~T.}\ \bibnamefont
  {Fofang}}, \bibinfo {author} {\bibfnamefont {T.-H.}\ \bibnamefont {Park}},
  \bibinfo {author} {\bibfnamefont {O.}~\bibnamefont {Neumann}}, \bibinfo
  {author} {\bibfnamefont {N.~A.}\ \bibnamefont {Mirin}}, \bibinfo {author}
  {\bibfnamefont {P.}~\bibnamefont {Nordlander}}, \ and\ \bibinfo {author}
  {\bibfnamefont {N.~J.}\ \bibnamefont {Halas}},\ }\href {\doibase
  10.1021/nl8024278} {\bibfield  {journal} {\bibinfo  {journal} {Nano Lett.}\
  }\textbf {\bibinfo {volume} {8}},\ \bibinfo {pages} {3481} (\bibinfo {year}
  {2008})}\BibitemShut {NoStop}%
\bibitem [{\citenamefont {T\"{o}rm\"{a}}\ and\ \citenamefont
  {Barnes}(2015)}]{torma_rpp78}%
  \BibitemOpen
  \bibfield  {author} {\bibinfo {author} {\bibfnamefont {P.}~\bibnamefont
  {T\"{o}rm\"{a}}}\ and\ \bibinfo {author} {\bibfnamefont {W.~L.}\ \bibnamefont
  {Barnes}},\ }\href {\doibase 10.1088/0034-4885/78/1/013901} {\bibfield
  {journal} {\bibinfo  {journal} {Rep. Prog. Phys.}\ }\textbf {\bibinfo
  {volume} {78}},\ \bibinfo {pages} {013901} (\bibinfo {year}
  {2015})}\BibitemShut {NoStop}%
\bibitem [{\citenamefont {Tserkezis}\ \emph
  {et~al.}(2018{\natexlab{a}})\citenamefont {Tserkezis}, \citenamefont
  {Gon\c{c}alves}, \citenamefont {Wolff}, \citenamefont {Todisco},
  \citenamefont {Busch},\ and\ \citenamefont {Mortensen}}]{tserkezis_prb98}%
  \BibitemOpen
  \bibfield  {author} {\bibinfo {author} {\bibfnamefont {C.}~\bibnamefont
  {Tserkezis}}, \bibinfo {author} {\bibfnamefont {P.~A.~D.}\ \bibnamefont
  {Gon\c{c}alves}}, \bibinfo {author} {\bibfnamefont {C.}~\bibnamefont
  {Wolff}}, \bibinfo {author} {\bibfnamefont {F.}~\bibnamefont {Todisco}},
  \bibinfo {author} {\bibfnamefont {K.}~\bibnamefont {Busch}}, \ and\ \bibinfo
  {author} {\bibfnamefont {N.~A.}\ \bibnamefont {Mortensen}},\ }\href {\doibase
  10.1103/PhysRevB.98.155439} {\bibfield  {journal} {\bibinfo  {journal} {Phys.
  Rev. B}\ }\textbf {\bibinfo {volume} {98}},\ \bibinfo {pages} {155439}
  (\bibinfo {year} {2018}{\natexlab{a}})}\BibitemShut {NoStop}%
\bibitem [{\citenamefont {Todisco}\ \emph {et~al.}(2020)\citenamefont
  {Todisco}, \citenamefont {Malureanu}, \citenamefont {Wolff}, \citenamefont
  {Gon\c{c}alves}, \citenamefont {Roberts}, \citenamefont {Mortensen},\ and\
  \citenamefont {Tserkezis}}]{todisco_nanoph9}%
  \BibitemOpen
  \bibfield  {author} {\bibinfo {author} {\bibfnamefont {F.}~\bibnamefont
  {Todisco}}, \bibinfo {author} {\bibfnamefont {R.}~\bibnamefont {Malureanu}},
  \bibinfo {author} {\bibfnamefont {C.}~\bibnamefont {Wolff}}, \bibinfo
  {author} {\bibfnamefont {P.~A.~D.}\ \bibnamefont {Gon\c{c}alves}}, \bibinfo
  {author} {\bibfnamefont {A.~S.}\ \bibnamefont {Roberts}}, \bibinfo {author}
  {\bibfnamefont {N.~A.}\ \bibnamefont {Mortensen}}, \ and\ \bibinfo {author}
  {\bibfnamefont {C.}~\bibnamefont {Tserkezis}},\ }\href {\doibase
  10.1515/nanoph-2019-0444} {\bibfield  {journal} {\bibinfo  {journal}
  {Nanophotonics}\ }\textbf {\bibinfo {volume} {9}},\ \bibinfo {pages} {803}
  (\bibinfo {year} {2020})}\BibitemShut {NoStop}%
\bibitem [{\citenamefont {Castellanos}\ \emph {et~al.}(2020)\citenamefont
  {Castellanos}, \citenamefont {Murai}, \citenamefont {Raziman}, \citenamefont
  {Wang}, \citenamefont {Ramezani}, \citenamefont {Curto},\ and\ \citenamefont
  {{G\'{o}mez Rivas}}}]{castellanos_acsphot7}%
  \BibitemOpen
  \bibfield  {author} {\bibinfo {author} {\bibfnamefont {G.~W.}\ \bibnamefont
  {Castellanos}}, \bibinfo {author} {\bibfnamefont {S.}~\bibnamefont {Murai}},
  \bibinfo {author} {\bibfnamefont {T.~V.}\ \bibnamefont {Raziman}}, \bibinfo
  {author} {\bibfnamefont {S.}~\bibnamefont {Wang}}, \bibinfo {author}
  {\bibfnamefont {M.}~\bibnamefont {Ramezani}}, \bibinfo {author}
  {\bibfnamefont {A.~G.}\ \bibnamefont {Curto}}, \ and\ \bibinfo {author}
  {\bibfnamefont {J.}~\bibnamefont {{G\'{o}mez Rivas}}},\ }\href {\doibase
  10.1021/acsphotonics.0c00063} {\bibfield  {journal} {\bibinfo  {journal} {ACS
  Photonics}\ }\textbf {\bibinfo {volume} {7}},\ \bibinfo {pages} {1226}
  (\bibinfo {year} {2020})}\BibitemShut {NoStop}%
\bibitem [{\citenamefont {Wu}\ \emph {et~al.}(2021)\citenamefont {Wu},
  \citenamefont {Guo}, \citenamefont {Huang}, \citenamefont {Liang},
  \citenamefont {Jin}, \citenamefont {Li}, \citenamefont {Deng}, \citenamefont
  {Jiao}, \citenamefont {Liu}, \citenamefont {Zhang}, \citenamefont {Zhang},\
  and\ \citenamefont {Yu}}]{wu_nn15}%
  \BibitemOpen
  \bibfield  {author} {\bibinfo {author} {\bibfnamefont {F.}~\bibnamefont
  {Wu}}, \bibinfo {author} {\bibfnamefont {J.}~\bibnamefont {Guo}}, \bibinfo
  {author} {\bibfnamefont {Y.}~\bibnamefont {Huang}}, \bibinfo {author}
  {\bibfnamefont {K.}~\bibnamefont {Liang}}, \bibinfo {author} {\bibfnamefont
  {L.}~\bibnamefont {Jin}}, \bibinfo {author} {\bibfnamefont {J.}~\bibnamefont
  {Li}}, \bibinfo {author} {\bibfnamefont {X.}~\bibnamefont {Deng}}, \bibinfo
  {author} {\bibfnamefont {R.}~\bibnamefont {Jiao}}, \bibinfo {author}
  {\bibfnamefont {Y.}~\bibnamefont {Liu}}, \bibinfo {author} {\bibfnamefont
  {J.}~\bibnamefont {Zhang}}, \bibinfo {author} {\bibfnamefont
  {W.}~\bibnamefont {Zhang}}, \ and\ \bibinfo {author} {\bibfnamefont
  {L.}~\bibnamefont {Yu}},\ }\href {\doibase 10.1021/acsnano.0c08274}
  {\bibfield  {journal} {\bibinfo  {journal} {ACS Nano}\ }\textbf {\bibinfo
  {volume} {15}},\ \bibinfo {pages} {2292} (\bibinfo {year}
  {2021})}\BibitemShut {NoStop}%
\bibitem [{\citenamefont {Zhu}\ \emph {et~al.}(2021)\citenamefont {Zhu},
  \citenamefont {Wu}, \citenamefont {Han}, \citenamefont {Shang}, \citenamefont
  {Liu}, \citenamefont {Yu}, \citenamefont {Yu}, \citenamefont {Li},\ and\
  \citenamefont {Ding}}]{zhu_nl21}%
  \BibitemOpen
  \bibfield  {author} {\bibinfo {author} {\bibfnamefont {J.}~\bibnamefont
  {Zhu}}, \bibinfo {author} {\bibfnamefont {F.}~\bibnamefont {Wu}}, \bibinfo
  {author} {\bibfnamefont {Z.}~\bibnamefont {Han}}, \bibinfo {author}
  {\bibfnamefont {Y.}~\bibnamefont {Shang}}, \bibinfo {author} {\bibfnamefont
  {F.}~\bibnamefont {Liu}}, \bibinfo {author} {\bibfnamefont {H.}~\bibnamefont
  {Yu}}, \bibinfo {author} {\bibfnamefont {L.}~\bibnamefont {Yu}}, \bibinfo
  {author} {\bibfnamefont {N.}~\bibnamefont {Li}}, \ and\ \bibinfo {author}
  {\bibfnamefont {B.}~\bibnamefont {Ding}},\ }\href {\doibase
  10.1021/acs.nanolett.1c00596} {\bibfield  {journal} {\bibinfo  {journal}
  {Nano Lett.}\ }\textbf {\bibinfo {volume} {21}},\ \bibinfo {pages} {3573}
  (\bibinfo {year} {2021})}\BibitemShut {NoStop}%
\bibitem [{\citenamefont {Condon}(1937)}]{condon_rmp9}%
  \BibitemOpen
  \bibfield  {author} {\bibinfo {author} {\bibfnamefont {E.~U.}\ \bibnamefont
  {Condon}},\ }\href {\doibase 10.1103/RevModPhys.9.432} {\bibfield  {journal}
  {\bibinfo  {journal} {Rev. Mod. Phys.}\ }\textbf {\bibinfo {volume} {9}},\
  \bibinfo {pages} {432} (\bibinfo {year} {1937})}\BibitemShut {NoStop}%
\bibitem [{\citenamefont {Tang}\ and\ \citenamefont
  {Cohen}(2010)}]{tang_prl104}%
  \BibitemOpen
  \bibfield  {author} {\bibinfo {author} {\bibfnamefont {Y.}~\bibnamefont
  {Tang}}\ and\ \bibinfo {author} {\bibfnamefont {A.~E.}\ \bibnamefont
  {Cohen}},\ }\href {\doibase 10.1103/PhysRevLett.104.163901} {\bibfield
  {journal} {\bibinfo  {journal} {Phys. Rev. Lett.}\ }\textbf {\bibinfo
  {volume} {104}},\ \bibinfo {pages} {163901} (\bibinfo {year}
  {2010})}\BibitemShut {NoStop}%
\bibitem [{\citenamefont {Fernandez-Corbaton}\ \emph
  {et~al.}(2013)\citenamefont {Fernandez-Corbaton}, \citenamefont
  {Zambrana-Puyalto}, \citenamefont {Tischler}, \citenamefont {Vidal},
  \citenamefont {Juan},\ and\ \citenamefont
  {Molina-Terriza}}]{fernandez_prl111}%
  \BibitemOpen
  \bibfield  {author} {\bibinfo {author} {\bibfnamefont {I.}~\bibnamefont
  {Fernandez-Corbaton}}, \bibinfo {author} {\bibfnamefont {X.}~\bibnamefont
  {Zambrana-Puyalto}}, \bibinfo {author} {\bibfnamefont {N.}~\bibnamefont
  {Tischler}}, \bibinfo {author} {\bibfnamefont {X.}~\bibnamefont {Vidal}},
  \bibinfo {author} {\bibfnamefont {M.~L.}\ \bibnamefont {Juan}}, \ and\
  \bibinfo {author} {\bibfnamefont {G.}~\bibnamefont {Molina-Terriza}},\ }\href
  {\doibase 10.1103/PhysRevLett.111.060401} {\bibfield  {journal} {\bibinfo
  {journal} {Phys. Rev. Lett.}\ }\textbf {\bibinfo {volume} {111}},\ \bibinfo
  {pages} {060401} (\bibinfo {year} {2013})}\BibitemShut {NoStop}%
\bibitem [{\citenamefont {Bohren}\ and\ \citenamefont
  {Huffman}(1983)}]{Bohren_Wiley1983}%
  \BibitemOpen
  \bibfield  {author} {\bibinfo {author} {\bibfnamefont {C.~F.}\ \bibnamefont
  {Bohren}}\ and\ \bibinfo {author} {\bibfnamefont {D.~R.}\ \bibnamefont
  {Huffman}},\ }\href {\doibase 10.1002/9783527618156} {\emph {\bibinfo {title}
  {Absorption and Scattering of Light by Small Particles}}}\ (\bibinfo
  {publisher} {John Wiley and Sons},\ \bibinfo {address} {New York},\ \bibinfo
  {year} {1983})\BibitemShut {NoStop}%
\bibitem [{\citenamefont {Mie}(1908)}]{mie_annphys330}%
  \BibitemOpen
  \bibfield  {author} {\bibinfo {author} {\bibfnamefont {G.}~\bibnamefont
  {Mie}},\ }\href {\doibase 10.1002/andp.19083300302} {\bibfield  {journal}
  {\bibinfo  {journal} {Ann. Phys.}\ }\textbf {\bibinfo {volume} {330}},\
  \bibinfo {pages} {377} (\bibinfo {year} {1908})}\BibitemShut {NoStop}%
\bibitem [{\citenamefont {Tserkezis}\ \emph {et~al.}(2020)\citenamefont
  {Tserkezis}, \citenamefont {Fern\'{a}ndez-Dom\'{i}nguez}, \citenamefont
  {Gon\c{c}alves}, \citenamefont {Todisco}, \citenamefont {Cox}, \citenamefont
  {Busch}, \citenamefont {Stenger}, \citenamefont {Bozhevolnyi}, \citenamefont
  {Mortensen},\ and\ \citenamefont {Wolff}}]{tserkezis_rpp83}%
  \BibitemOpen
  \bibfield  {author} {\bibinfo {author} {\bibfnamefont {C.}~\bibnamefont
  {Tserkezis}}, \bibinfo {author} {\bibfnamefont {A.~I.}\ \bibnamefont
  {Fern\'{a}ndez-Dom\'{i}nguez}}, \bibinfo {author} {\bibfnamefont {P.~A.~D.}\
  \bibnamefont {Gon\c{c}alves}}, \bibinfo {author} {\bibfnamefont
  {F.}~\bibnamefont {Todisco}}, \bibinfo {author} {\bibfnamefont {J.~D.}\
  \bibnamefont {Cox}}, \bibinfo {author} {\bibfnamefont {K.}~\bibnamefont
  {Busch}}, \bibinfo {author} {\bibfnamefont {N.}~\bibnamefont {Stenger}},
  \bibinfo {author} {\bibfnamefont {S.~I.}\ \bibnamefont {Bozhevolnyi}},
  \bibinfo {author} {\bibfnamefont {N.~A.}\ \bibnamefont {Mortensen}}, \ and\
  \bibinfo {author} {\bibfnamefont {C.}~\bibnamefont {Wolff}},\ }\href
  {\doibase 10.1088/1361-6633/aba348} {\bibfield  {journal} {\bibinfo
  {journal} {Rep. Prog. Phys.}\ }\textbf {\bibinfo {volume} {83}},\ \bibinfo
  {pages} {082401} (\bibinfo {year} {2020})}\BibitemShut {NoStop}%
\bibitem [{\citenamefont {Antosiewicz}\ \emph {et~al.}(2014)\citenamefont
  {Antosiewicz}, \citenamefont {Apell},\ and\ \citenamefont
  {Shegai}}]{antosiewicz_acsphot1}%
  \BibitemOpen
  \bibfield  {author} {\bibinfo {author} {\bibfnamefont {T.~J.}\ \bibnamefont
  {Antosiewicz}}, \bibinfo {author} {\bibfnamefont {S.~P.}\ \bibnamefont
  {Apell}}, \ and\ \bibinfo {author} {\bibfnamefont {T.}~\bibnamefont
  {Shegai}},\ }\href {\doibase 10.1021/ph500032d} {\bibfield  {journal}
  {\bibinfo  {journal} {ACS Photonics}\ }\textbf {\bibinfo {volume} {1}},\
  \bibinfo {pages} {454} (\bibinfo {year} {2014})}\BibitemShut {NoStop}%
\bibitem [{\citenamefont {Fan}\ and\ \citenamefont
  {Govorov}(2011)}]{fan_jpcc115}%
  \BibitemOpen
  \bibfield  {author} {\bibinfo {author} {\bibfnamefont {Z.}~\bibnamefont
  {Fan}}\ and\ \bibinfo {author} {\bibfnamefont {A.~O.}\ \bibnamefont
  {Govorov}},\ }\href {\doibase 10.1021/jp204265x} {\bibfield  {journal}
  {\bibinfo  {journal} {J. Phys. Chem. C}\ }\textbf {\bibinfo {volume} {115}},\
  \bibinfo {pages} {13254} (\bibinfo {year} {2011})}\BibitemShut {NoStop}%
\bibitem [{\citenamefont {Song}\ \emph {et~al.}(2013)\citenamefont {Song},
  \citenamefont {Blaber}, \citenamefont {Zhao}, \citenamefont {Zhang},
  \citenamefont {Fry}, \citenamefont {Schatz},\ and\ \citenamefont
  {Rosi}}]{song_nl13}%
  \BibitemOpen
  \bibfield  {author} {\bibinfo {author} {\bibfnamefont {C.}~\bibnamefont
  {Song}}, \bibinfo {author} {\bibfnamefont {M.~G.}\ \bibnamefont {Blaber}},
  \bibinfo {author} {\bibfnamefont {G.}~\bibnamefont {Zhao}}, \bibinfo {author}
  {\bibfnamefont {P.}~\bibnamefont {Zhang}}, \bibinfo {author} {\bibfnamefont
  {H.~C.}\ \bibnamefont {Fry}}, \bibinfo {author} {\bibfnamefont {G.~C.}\
  \bibnamefont {Schatz}}, \ and\ \bibinfo {author} {\bibfnamefont {N.~L.}\
  \bibnamefont {Rosi}},\ }\href {\doibase 10.1021/nl4013776} {\bibfield
  {journal} {\bibinfo  {journal} {Nano Lett.}\ }\textbf {\bibinfo {volume}
  {13}},\ \bibinfo {pages} {3256} (\bibinfo {year} {2013})}\BibitemShut
  {NoStop}%
\bibitem [{\citenamefont {Kuzyk}\ \emph {et~al.}(2018)\citenamefont {Kuzyk},
  \citenamefont {Jungmann}, \citenamefont {Acuna},\ and\ \citenamefont
  {Liu}}]{kuzyk_acsphot5}%
  \BibitemOpen
  \bibfield  {author} {\bibinfo {author} {\bibfnamefont {A.}~\bibnamefont
  {Kuzyk}}, \bibinfo {author} {\bibfnamefont {R.}~\bibnamefont {Jungmann}},
  \bibinfo {author} {\bibfnamefont {G.~P.}\ \bibnamefont {Acuna}}, \ and\
  \bibinfo {author} {\bibfnamefont {N.}~\bibnamefont {Liu}},\ }\href {\doibase
  10.1021/acsphotonics.7b01580} {\bibfield  {journal} {\bibinfo  {journal} {ACS
  Photonics}\ }\textbf {\bibinfo {volume} {5}},\ \bibinfo {pages} {1151}
  (\bibinfo {year} {2018})}\BibitemShut {NoStop}%
\bibitem [{\citenamefont {Ye\c{s}ilyurt}\ and\ \citenamefont
  {Huang}(2021)}]{yesilyurt_aom9}%
  \BibitemOpen
  \bibfield  {author} {\bibinfo {author} {\bibfnamefont {A.~T.~M.}\
  \bibnamefont {Ye\c{s}ilyurt}}\ and\ \bibinfo {author} {\bibfnamefont {J.~S.}\
  \bibnamefont {Huang}},\ }\href {\doibase 10.1002/adom.202100848} {\bibfield
  {journal} {\bibinfo  {journal} {Adv. Opt. Mater.}\ }\textbf {\bibinfo
  {volume} {9}},\ \bibinfo {pages} {2100848} (\bibinfo {year}
  {2021})}\BibitemShut {NoStop}%
\bibitem [{\citenamefont {Tserkezis}\ \emph
  {et~al.}(2018{\natexlab{b}})\citenamefont {Tserkezis}, \citenamefont
  {Ye\c{s}ilyurt}, \citenamefont {Huang},\ and\ \citenamefont
  {Mortensen}}]{tserkezis_acsphot5b}%
  \BibitemOpen
  \bibfield  {author} {\bibinfo {author} {\bibfnamefont {C.}~\bibnamefont
  {Tserkezis}}, \bibinfo {author} {\bibfnamefont {A.~T.~M.}\ \bibnamefont
  {Ye\c{s}ilyurt}}, \bibinfo {author} {\bibfnamefont {J.-S.}\ \bibnamefont
  {Huang}}, \ and\ \bibinfo {author} {\bibfnamefont {N.~A.}\ \bibnamefont
  {Mortensen}},\ }\href {\doibase 10.1021/acsphotonics.8b01261} {\bibfield
  {journal} {\bibinfo  {journal} {ACS Photonics}\ }\textbf {\bibinfo {volume}
  {5}},\ \bibinfo {pages} {5017} (\bibinfo {year}
  {2018}{\natexlab{b}})}\BibitemShut {NoStop}%
\bibitem [{\citenamefont {Johnson}\ and\ \citenamefont
  {Christy}(1972)}]{johnson_prb6}%
  \BibitemOpen
  \bibfield  {author} {\bibinfo {author} {\bibfnamefont {P.~B.}\ \bibnamefont
  {Johnson}}\ and\ \bibinfo {author} {\bibfnamefont {R.~W.}\ \bibnamefont
  {Christy}},\ }\href {\doibase 10.1103/PhysRevB.6.4370} {\bibfield  {journal}
  {\bibinfo  {journal} {Phys. Rev. B}\ }\textbf {\bibinfo {volume} {6}},\
  \bibinfo {pages} {4370} (\bibinfo {year} {1972})}\BibitemShut {NoStop}%
\bibitem [{\citenamefont {Liu}\ and\ \citenamefont
  {Liedl}(2018)}]{liu_chemrev118}%
  \BibitemOpen
  \bibfield  {author} {\bibinfo {author} {\bibfnamefont {N.}~\bibnamefont
  {Liu}}\ and\ \bibinfo {author} {\bibfnamefont {T.}~\bibnamefont {Liedl}},\
  }\href {\doibase 10.1021/acs.chemrev.7b00225} {\bibfield  {journal} {\bibinfo
   {journal} {Chem. Rev.}\ }\textbf {\bibinfo {volume} {118}},\ \bibinfo
  {pages} {3032} (\bibinfo {year} {2018})}\BibitemShut {NoStop}%
\bibitem [{\citenamefont {Tserkezis}\ \emph {et~al.}(2014)\citenamefont
  {Tserkezis}, \citenamefont {Taylor}, \citenamefont {Beitner}, \citenamefont
  {Esteban}, \citenamefont {Baumberg},\ and\ \citenamefont
  {Aizpurua}}]{tserkezis_part31}%
  \BibitemOpen
  \bibfield  {author} {\bibinfo {author} {\bibfnamefont {C.}~\bibnamefont
  {Tserkezis}}, \bibinfo {author} {\bibfnamefont {R.~W.}\ \bibnamefont
  {Taylor}}, \bibinfo {author} {\bibfnamefont {J.}~\bibnamefont {Beitner}},
  \bibinfo {author} {\bibfnamefont {R.}~\bibnamefont {Esteban}}, \bibinfo
  {author} {\bibfnamefont {J.~J.}\ \bibnamefont {Baumberg}}, \ and\ \bibinfo
  {author} {\bibfnamefont {J.}~\bibnamefont {Aizpurua}},\ }\href {\doibase
  10.1002/ppsc.201300287} {\bibfield  {journal} {\bibinfo  {journal} {Part.
  Part. Syst. Charact.}\ }\textbf {\bibinfo {volume} {31}},\ \bibinfo {pages}
  {152} (\bibinfo {year} {2014})}\BibitemShut {NoStop}%
\end{thebibliography}%

\end{document}